\documentclass[12pt,a4paper]{article}
\usepackage[format=hang]{caption}
\usepackage{epsfig,amssymb,amsmath,graphicx,subcaption,verbatim,hyperref,xcolor,ulem,sidecap,bm,float}
\newcommand*\rfrac[2]{{}^{#1}\!/_{#2}}
\newcommand{\be}{\begin{equation}}
\newcommand{\ee}{\end{equation}}

\newcommand{\bpi}{\mbox{\boldmath $\pi$}}
\newcommand{\pauli}{\mbox{\boldmath $\tau$}}

\DeclareMathOperator{\Tr}{Tr}

\newcommand{\news}{\setcounter{equation}{0}}
\def\bea{\begin{eqnarray}}
\def\eea{\end{eqnarray}}
\numberwithin{equation}{section}
\makeatletter
\renewcommand*\env@matrix[1][\arraystretch]{
  \edef\arraystretch{#1}
  \hskip -\arraycolsep
  \let\@ifnextchar\new@ifnextchar
  \array{*\c@MaxMatrixCols c}}
\makeatother

\begin{document} 
\title{\vskip -65pt
\begin{flushright}
{\normalsize DAMTP-2014-13} \\
\end{flushright}
\vskip 60pt
{\bf {\LARGE Quantization of $T_d$- and $O_h$-symmetric Skyrmions}}\\[20pt]}
\author{\bf {\Large P.H.C. Lau\footnote{P.H.C.Lau@damtp.cam.ac.uk} \ and N.S. Manton\footnote{N.S.Manton@damtp.cam.ac.uk}} \\[25pt]
Department of Applied Mathematics and Theoretical Physics\\
University of Cambridge\\
Wilberforce Road, Cambridge CB3 0WA, UK}

\date{February 2014}
\maketitle
\vskip 20pt
 
\begin{abstract}

The geometrical construction of rational maps using a cubic grid has led to many new Skyrmion solutions, with baryon numbers up to 108. Energy spectra of some of the new Skyrmions are calculated here by semi-classical quantization. Quantization of the $B=20$ $T_d$-symmetric Skyrmion, which is one of the newly found Skyrmions, is considered, and this leads to the development of a new approach to solving Finkelstein-Rubinstein (F-R) constraints. Matrix equations are simplified by introducing a Cartesian version of angular momentum basis states, and the computations are easier. The quantum states of all $T_d$-symmetric Skyrmions, constructed from the cubic grid, are classified into three classes, depending on the contribution of vertex points of the cubic grid to the rational maps. The analysis is extended to the larger symmetry group $O_h$. Quantum states of $O_h$-symmetric Skyrmions, constructed from the cubic grid, form a subset of the $T_d$-symmetric quantum states.
\end{abstract}

\vskip 20pt

\newpage

\section{Introduction}\news

The Skyrme model is a soliton model for nuclear physics \cite{Sk,book,BR}. The soliton solutions are called Skyrmions, and the conserved integer-valued topological charge of the Skyrmions is interpreted as the baryon number of nuclei. The Skyrme Lagrangian is
\be
L= \int \left\{ -\frac{F^2_\pi}{16} \Tr(R_{\mu}R^{\mu}) + \frac{1}{32e^2} \Tr([R_{\mu},R_{\nu}][R^{\mu},R^{\nu}]) + \frac{m^2_\pi F^2_\pi}{8} \Tr(U-{\boldsymbol I}) \right\} \, d^3x \, ,
\label{eqn:SkyLag}
\ee
\noindent where $U(x)$ is the $SU(2)$-valued scalar Skyrme field, $R_{\mu} = \partial _{\mu} U U^{\dagger}$ is the right current, and $m_\pi$ is the pion mass. The free parameters of the model ($F_\pi$, $e$, $m_\pi$) can be absorbed by using the Skyrme mass unit $\frac{F_\pi}{4e}$ and length unit $\frac{2}{eF_\pi}$. The scaled pion mass is $\frac{2m_\pi}{eF_\pi}$, and in this paper, a value of unity is used. The Skyrme field $U$ can be written in terms of the pion fields $\bpi = (\pi_1, \pi_2, \pi_3)$ and sigma field $\sigma$ as
\be
U(x) = \sigma(x) \, {\boldsymbol I} + i \bpi(x) \cdot \pauli \,,
\ee
where $\pauli$ are the Pauli matrices. The Skyrme field is subject to the constraint $UU^{\dagger} = (\sigma^2 + \bpi \cdot \bpi){\boldsymbol I} = {\boldsymbol I}$. Hence, $\sigma^2 + \bpi \cdot \bpi =1$ and $\sigma$ and $\bpi$ are not independent.

The baryon number, the topological charge of a Skyrmion, is the integral of the baryon density
\be
{\cal B} = -\frac{1}{24\pi^2}\epsilon_{ijk}{\rm Tr}(R_iR_jR_k) \,. 
\label{eq:bardens}
\ee
The angular structure of Skyrmions is not known precisely except for the $B=1$ ``hedgehog''. For others with higher baryon number, solutions can be found numerically with the help of the rational map ansatz \cite{HMS}. An initial field configuration is constructed from a rational map, which encodes the angular structure, and a radial profile function, and it is then numerically relaxed to the stable Skyrmion. A geometrical construction of rational maps using a cubic grid was developed in \cite{FLM}, and a range of new Skyrmions was found. Most of the new solutions retain the symmetry of the rational maps, which are subgroups of the cubic symmetry group $O_h$.

The $T_d$- and $O_h$-symmetric Skyrmions are of particular interest to us, because of various theoretical studies of possibly tetrahedral and cubic nuclei \cite{DGSM,DCDDPOS}. The idea of tetrahedrally-symmetric nuclei is supported by some experimental results. The characteristic $3^-$ state, which is the first allowed rotational excitation of a tetrahedral rigid body with $0^+$ ground state, can be seen experimentally in the spectra of $^{16}$O and $^{40}$Ca \cite{TWC,CS}. The next excited state is a $4^+$ state.

The rotational spectra of Skyrmions can be calculated using semi-classical quantization \cite{ANW,BC}. The spectra of isospin excitations can also be calculated. A tetrahedrally-symmetric $B=16$ Skyrmion is known \cite{BMS} and several others have been found recently. The $3^-$ quantum state together with the second rotational state, $4^+$, which we will calculate using the Cartesian method in Section 4, can be used to identify potential $T_d$-symmetric nuclei from experiments. The ratio of the excitation energies between two rotational states with spins $l_2$ and $l_1$, respectively, is $\frac{l_2(l_2+1)}{l_1(l_1+1)}$. This equals $\frac{5}{3}\sim1.67$ for the $4^+$ and $3^-$ states. The existence of these two states as the lowest rotational excitations with a ratio of energies approximately $\frac{5}{3}$ suggests that a nucleus is tetrahedrally-symmetric. Such states can be seen clearly in $^{16}$O. The energy of the lowest $3^-$ and $4^+$ states of $^{16}$O are $6.130$ and $10.356$ MeV, respectively, with a ratio of $1.69$ \cite{TWC}. The procedure is reviewed and applied to the $B=20$ $T_d$-symmetric Skyrmion below. We show more generally that the quantum states of $T_d$-symmetric Skyrmions constructed from the cubic grid, which all have baryon number a multiple of 4, are classified into three classes. 

In Section 2, we briefly review semi-classical Skyrmion quantization. In Section 3, we recall the cubic grid method, based on the Skyrme crystal structure, for
constructing rational maps. The $B=20$ $T_d$-symmetric Skyrmion and its rational map are described here. Section 4 discusses the semi-classical quantization of this Skyrmion. We introduce a method for writing down quantum states using Cartesian coordinates rather than Euler angles. This simplifies the analysis of states with $T_d$ and $O_h$ symmetry. Energy levels of the $B=20$ $T_d$-symmetric Skyrmions are calculated and presented. We also describe the use of the cubic grid to construct the parity operator pictorially. Sections 5 and 6 discuss the extension of the Cartesian method to a large class of Skyrmions with $T_d$ and $O_h$ symmetries, respectively. Concluding remarks are in Section 7, and our numerical methods are discussed in the Appendix.

\section{Semi-classical Skyrmion quantization}
In the semi-classical quantization method, a Skyrmion is treated as a rigid body free to rotate in ordinary space and isospace. The Skyrmion is restricted to have rotational and isorotational degrees of freedom only \cite{ANW,BC} and is parametrized as
\be
U( \mathbf{x} , \mathbf{A}, \mathbf{B})  \equiv \mathbf{A} U_0(\mathbf{R}(\mathbf{B})\mathbf{x})\mathbf{A}^{\dagger} \, .
\label{eqn:U_rot}
\ee
$U_0(\mathbf{x})$ is the prepared static Skyrmion solution in some convenient fixed orientation, $\mathbf{A}$ is the $SU(2)$ isospatial rotation, and $\mathbf{R(B)}$ is the spatial rotation represented by an $SU(2)$ matrix $\mathbf{B}$. Classically, $\mathbf{A}$ and $\mathbf{B}$ are time-dependent.

The semi-classical quantization of Skyrmions generalizes the quantization of a rigid body of normal matter. The classical rotational energy of an isotropic rigid body is
\be
E = \frac{L^{2}}{2I} \, ,
\ee
where $I$ is the moment of inertia of the body. To quantize the rigid body, we promote the squared angular momentum $L^2$ to a quantum operator $\hat{L}^2$. The eigenvalue of $\hat{L}^2$ is $L(L+1)$ in a state of angular momentum $L$. This formalism can be applied to Skyrmions, but the situation is more complicated. Skyrmions possess both spin and isospin, and the inertia tensors are not generally isotropic.

Substituting (\ref{eqn:U_rot}) into the Skyrme Lagrangian (\ref{eqn:SkyLag}), one can show after some rearrangements that the kinetic part of the Lagrangian is
\be
H_{\textrm{kin}} = {\frac{1}{2}} a_{i} U_{ij} a_{j} - a_{i} W_{ij} b_{j} + {\frac{1}{2}} b_{i} V_{ij} b_{j} \, ,
\label{eqn:kin}
\ee
\noindent where $a_j$ and $b_j$ are the angular velocities in isospace and ordinary space defined by
\be
a_{j} = - i \Tr \left( \pauli _{j} A ^{\dagger} \dot{A}\right) \, , \quad b_{j} = i \Tr \left( \pauli _{j} \dot{B} B^{\dagger}\right) \, .
\ee
\noindent $U_{ij}$, $W_{ij}$ and $V_{ij}$ are the moment of inertia tensors, which can be written in terms of the Skyrme field $U_0$, its right current $R_i$, and the further current $T_{i} = \frac{i}{2} [\pauli _i , U_0] U_0^{\dagger}$ as \cite{BC}
\begin{align}
U_{ij} &= - \int \, \Tr \left( T_i T_j + \frac{1}{4} [R_{k} , T_{i}][R_{k} , T_{j}] \right) \, d^{3}x \, , \label{eqn:U} \\
V_{ij} &= - \int \, \epsilon _{ilm} \epsilon _{jnp} x_{l} x_{n} \Tr \left( R_{m} R_{p} + \frac{1}{4} [R_{k} , R_{m}] [R_{k} , R_{p}] \right) \, d^{3}x \, ,  \label{eqn:V} \\
W_{ij} &= \int \, \epsilon _{jlm} x_{l} \Tr \left( T_{i} R_{m} + \frac{1}{4} [R_{k},T_{i}][R_{k} , R_{m}] \right) \, d^{3}x \, . \label{eqn:W}
\end{align}
\indent In order to calculate the nuclear spectra from the Skyrme model, one must first calculate numerically all the moments of inertia ($U_{ij}, W_{ij}, V_{ij}$) of a given Skyrmion. The details of the calculations can be found in the Appendix.

Skyrmions can be in any orientation with respect to the underlying coordinate system. One can introduce the body-fixed isospin ($\mathbf{K}$) and spin ($\mathbf{L}$), and the space-fixed isospin ($\mathbf{I}$) and spin ($\mathbf{J}$). The body-fixed isospin and spin are the conjugate momenta to the isospatial rotation $\mathbf{A}$ and the spatial rotation $\mathbf{B}$, derived from $H_{\textrm{kin}}$; the explicit relations in terms of the angular velocities $a_j$ and $b_j$ are
\be
K_i = U_{ij} a_{j} - W_{ij} b_{j} \,, \quad L_{i} = - W^{T}_{ij} a_{j} + V_{ij} b_{j} \, .
\label{eqn:ang}
\ee
The space-fixed isospin and spin can be obtained from the body-fixed isospin and spin by suitable rotations
\be
I_{i} = - R_{ij} (\mathbf{A}) K_{j} \,, \quad J_{i} = -R_{ij} (\mathbf{B})^{T} L_{j} \, .
\ee
\indent The four sets of quantized angular momentum operators mutually commute and satisfy the usual $SU(2)$ commutation relations
\begin{align}
[\hat{I}_{i},\hat{I}_{j}] &= i \epsilon _{ijk} \hat{I}_{k} \,, \quad \ \ [\hat{J}_{i},\hat{J}_{j}] = i \epsilon _{ijk} \hat{J}_{k} \, , \\
[\hat{K}_{i},\hat{K}_{j}] &= i \epsilon _{ijk} \hat{K}_{k} \,, \quad [\hat{L}_{i},\hat{L}_{j}] = i \epsilon _{ijk} \hat{L}_{k} \, .
\end{align}
The total angular momentum operators $\hat{I}^2$, $\hat{J}^2$, $\hat{K}^2$ and $\hat{L}^2$ are independent of orientations. As a result, the total isospin and total spin are equal in the body-fixed and space-fixed frames,
\be
\hat{I}^2 = \hat{K}^2 \,, \quad \hat{J}^2 = \hat{L}^2 \, .
\ee
\indent The quantized kinetic Hamiltonian of the Skyrmion can be written in the standard way in terms of the body-fixed angular momenta, $\hat{K}_i$ and $\hat{L}_i$. The Hamiltonian of a general Skyrmion with no symmetry is complicated, because of the cross term $W_{ij}$, which mixes isospin and spin. The expression is simple for symmetric Skyrmions. For example, the Hamiltonian of the $B=20$ $T_d$-symmetric Skyrmion is
\be
\hat{H}=\frac{1}{2v}\hat{L}^2 + \frac{1}{2U_{11}}(\hat{K}^2-\hat{K}_3^2) +\frac{1}{2U_{33}}\hat{K}_3^2 \,.
\label{eqn:TdHam}
\ee
\noindent Because of the $T_d$ symmetry, $V_{ij}=v \delta_{ij}$, $U_{11}=U_{22}$ and the isospin and spin contributions decouple because the cross term $W_{ij}$ vanishes. The symmetry also puts constraints on the spin and isospin quantum numbers. These are called Finkelstein-Rubinstein constraints \cite{FR}.

The wavefunction of the Skyrmion can be expressed as a tensor product of spin and isospin Wigner $D$-functions, $D^L_{J_3, L_3}(\phi, \theta, \psi) \otimes D^K_{I_3, K_3}(\alpha, \beta, \gamma)$ (or in quantum state notation $\left | L, L_3, J_3 \right \rangle \otimes \left | K, K_3, I_3 \right \rangle$), where $J_3$ and $I_3$ take all values in the standard range. The energy eigenvalues do not depend on the values of $J_3$ and $I_3$. For Skyrmions of even baryon number (which they all are in this paper), the spin and isospin are integer-valued. We can then simplify the quantum state by setting $J_3=I_3=0$. A $U(1)$ subgroup of the $SU(2)$ group is thereby quotiented out, and the manifold where the $D$-functions live is $S^3/S^1 \sim S^2$. The 2-sphere is parametrized by the remaining two angles, and the Wigner functions can be expressed in terms of the more familiar spherical harmonics. The product of Wigner functions $D^L_{0, L_3}(\phi,\theta,\psi) \otimes D^K_{0, K_3}(\alpha,\beta,\gamma)$ is proportional to $(-1)^{L_3}Y_{L, L_3}^*(\theta,\psi) \otimes (-1)^{K_3}Y_{K, K_3}^*(\beta,\gamma)$; the quantum state simplifies to $\left | L, L_3 \right \rangle \otimes \left | K, K_3 \right \rangle$.

The quantum state $\left | L, L_3 \right \rangle \otimes \left | K, K_3 \right \rangle$ is an eigenstate of the Hamiltonian operator (\ref{eqn:TdHam}). The energy of the state can be easily calculated, by replacing the operators $\hat{L}^2$, $\hat{K}^2$ and $\hat{K}_3$ with their respective eigenvalues $L(L+1)$, $K(K+1)$ and $K_3$. This energy is in Skyrme units; we can convert it back to physical units by a conversion factor. The rotational energy of the Skyrmion has the form $\frac{\hat{L}^2}{2v}$, where $\hat{L}^2$ is proportional to $\hbar^2$, but we have set $\hbar=1$. The moment of inertia $v$ has dimension $\text{[Mass][Length]}^2$. By inserting the Skyrme mass and length units, one Skyrme unit of moment of inertia is converted to $(\frac{F_\pi}{4e})(\frac{2}{eF_\pi})^2 = \frac{1}{e^3F_\pi}$ in physical units, which corresponds to an energy conversion factor $e^3F_\pi$ with a typical value of $O(10^3)$ MeV \cite{BMSW}. In this paper, a value $e^3F_\pi = 4 \times 10^3$ MeV is used.

\section{Rational maps from the cubic grid}\news

\begin{figure}[ht]
\centering
\hspace{4cm}
\includegraphics[width=8.5cm]{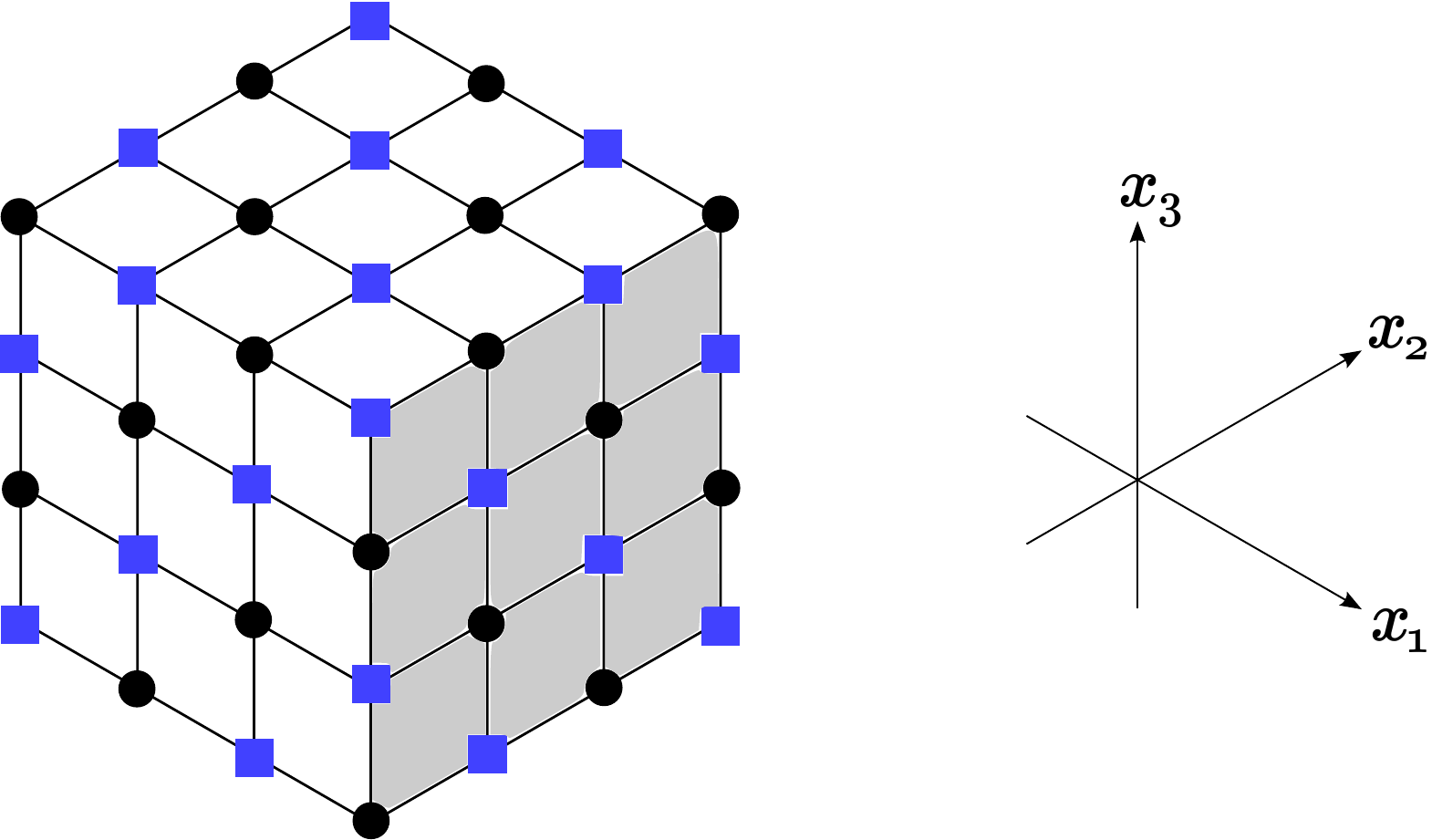}
\caption{Second layer of the cubic grid}
\label{fig:cube_grid}
\end{figure}

Finding rational maps with $T_d$ and $O_h$ symmetry becomes difficult as the degree of the maps becomes large. The cubic grid method for constructing them was first proposed in \cite{MSkySD} and developed in detail in \cite{FLM}. It relies on the observation that $\hat{\bpi}_3$ takes the values $\pm 1$ at the lattice points of the Skyrme crystal, corresponding to the zeros or poles of the rational map. A cubic chunk of the Skyrme crystal can be visually simplified to a cubic grid, and rational maps can be constructed by picking specific points from the grid. In general, large baryon number Skyrmions are found using a multi-layer rational map ansatz \cite{MP}. In the cubic grid method, the $n$-th layer rational map is constructed from a $2n \times 2n \times 2n$ grid. The first three layers of the grid allow a maximum number of 8, 56 and 152 points, respectively, and lead to rational maps of maximum degree 4, 28 and 76 if one restricts to zeros and poles of multiplicity one. The degree of the rational map is the baryon number of the layer. The example in Figure \ref{fig:cube_grid} shows the cubic grid method applied to the outer layer of the $B=32$ Skyrmion, which is the second layer of the cubic grid. The circles and squares indicate the 28 zeros and poles of the rational map, respectively. By labelling points with scaled Cartesian coordinates, ($x_1, \,x_2,\,x_3$), the rational map can be expressed in terms of the complex Riemann sphere coordinate, $\textrm{z}=\frac{x_1+ix_2}{r+x_3}$, with $r^2=x_1^2+x_2^2+x_3^2$. The locations of the zeros and poles can be found with the same formula.

\begin{figure}[h]
\centering
        \begin{subfigure}[b]{0.28\textwidth}
                \centering
		\includegraphics[width=0.95\textwidth]{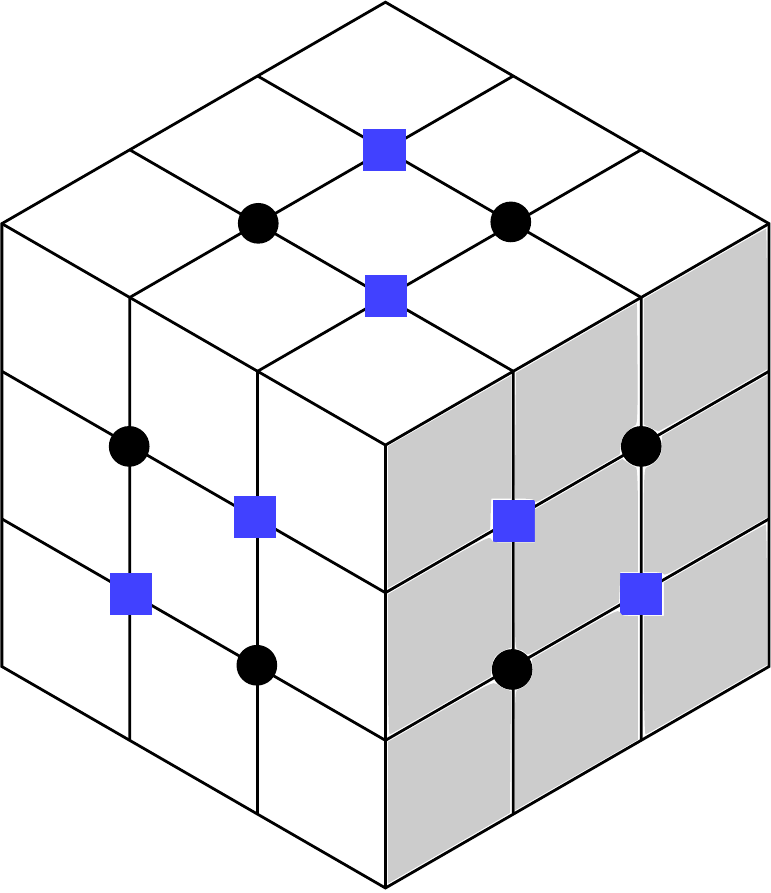}
                \caption{Points on face interiors}
                \label{fig:R_F}
        \end{subfigure}
        ~ \, 
        \begin{subfigure}[b]{0.28\textwidth}
                \centering
                \includegraphics[width=\textwidth]{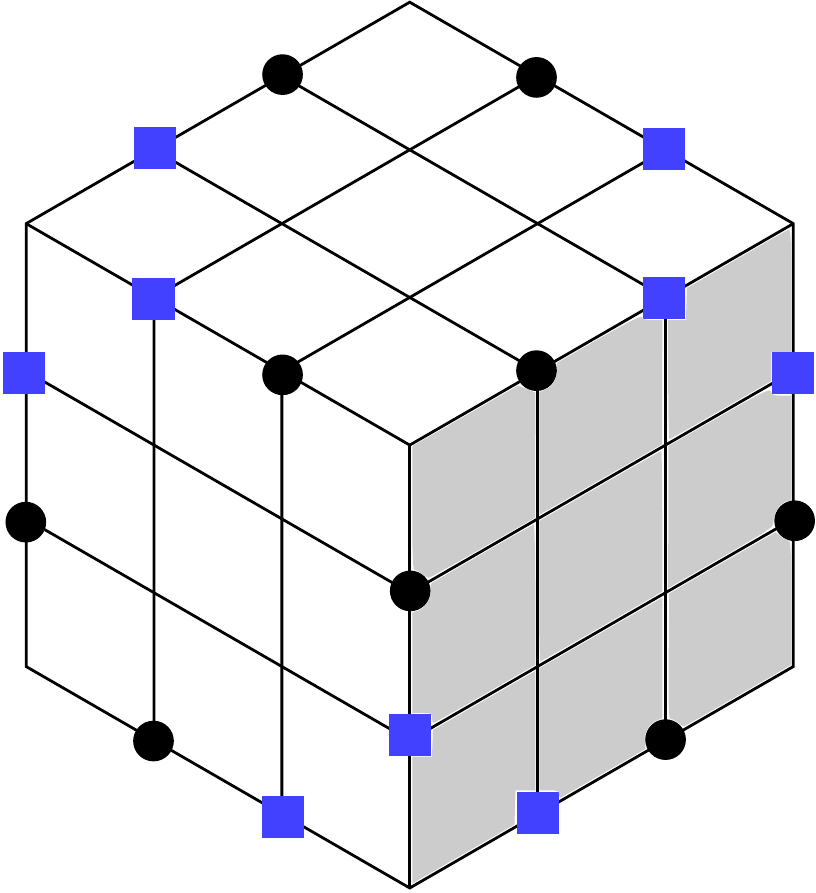}
                \caption{Points on edges}
                \label{fig:R_E}
        \end{subfigure}
        ~
        \begin{subfigure}[b]{0.28\textwidth}
                \centering
                \includegraphics[width=\textwidth]{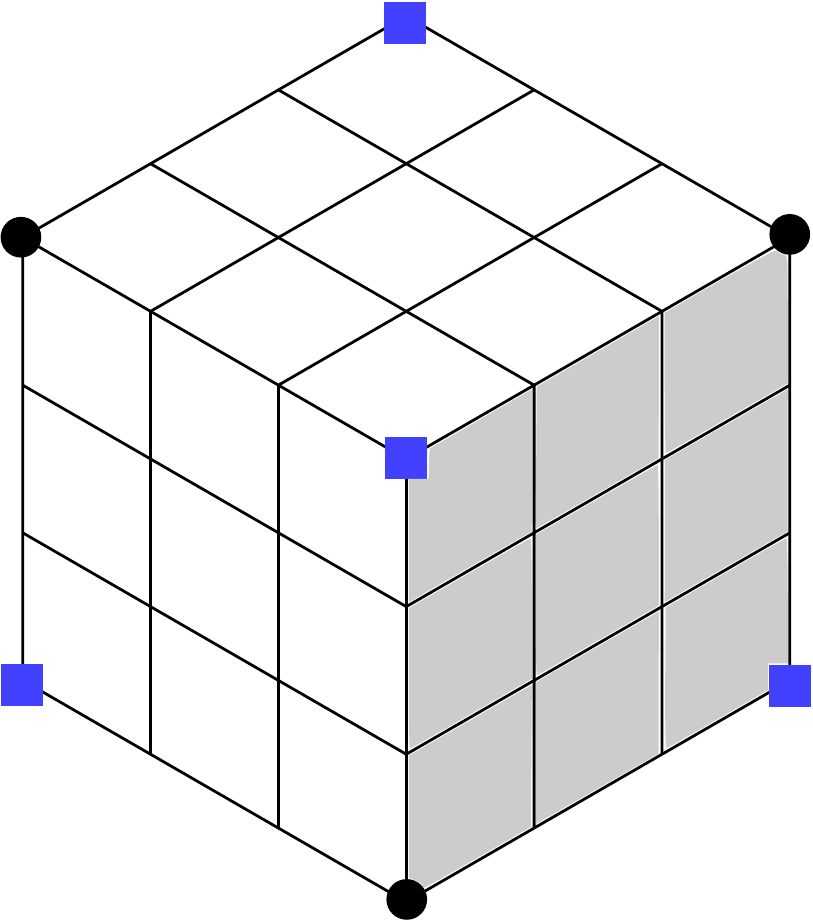}
                \caption{Points on vertices}
                \label{fig:R_v}
        \end{subfigure}
        \caption{Subsets of the $4 \times 4 \times 4$ cubic grid}
        \label{fig:grid}
\end{figure}

With the aid of the cubic grid, the symmetry of the rational map can be easily visualized too. The $B=32$ Skyrmion possesses cubic symmetry $O_h$, which is the full symmetry group of the grid. The points on the grid can be separated into smaller subsets of points still preserving the $O_h$ symmetry (see Figure \ref{fig:grid}).
\noindent $O_h$-symmetric rational maps of various degrees can be constructed by taking combinations of the subsets. One can relax the symmetry requirement, and obtain many further rational maps.

\begin{figure}[ht]
\centering
\includegraphics[width=0.3\textwidth]{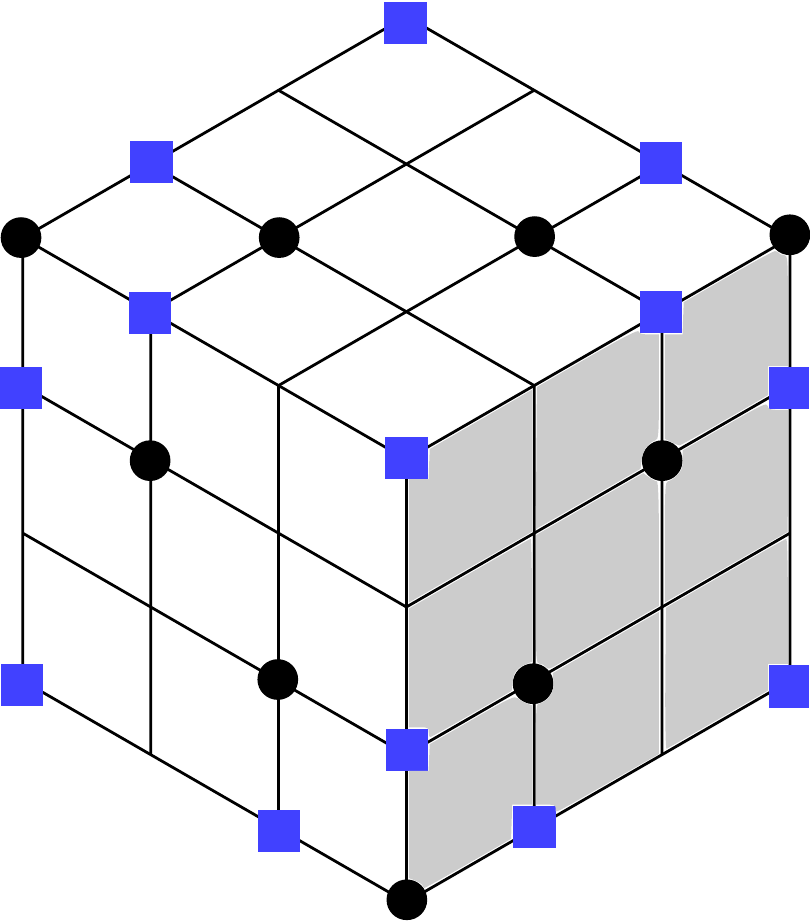}
\caption{Outer rational map for the $B=20$ $T_d$-symmetric Skyrmion}
\label{fig:B_20}
\end{figure}

The $B=20$ $T_d$-symmetric Skyrmion is found by using a double-layer rational map ansatz. The inner degree 4 rational map is constructed using all eight vertices of the first layer of the cubic grid. It is expressed in terms of the quartic Klein polynomials with zeros on alternating vertices, denoted by $p_+$ and $p_-$, respectively; these are

\begin{align}
p_+(\mathrm{z}) &= \left(\mathrm{z} + \frac{1-i}{\sqrt{3} + 1}\right) \left(\mathrm{z} - \frac{1-i}{\sqrt{3} + 1}\right) \left(\mathrm{z} + \frac{1+i}{\sqrt{3} - 1}\right) \left(\mathrm{z} - \frac{1+i}{\sqrt{3} - 1}\right) = \mathrm{z}^4 + 2 \sqrt{3}i \mathrm{z}^2 + 1 \,, \\
p_-(\mathrm{z}) &= \left(\mathrm{z} + \frac{1+i}{\sqrt{3} + 1}\right) \left(\mathrm{z} - \frac{1+i}{\sqrt{3} + 1}\right) \left(\mathrm{z} + \frac{1-i}{\sqrt{3} - 1}\right) \left(\mathrm{z} - \frac{1-i}{\sqrt{3} - 1}\right) = \mathrm{z}^4 - 2 \sqrt{3}i \mathrm{z}^2 + 1 \,.
\label{eqn:Klein}
\end{align}
The inner map is simply $\rfrac{p_+}{p_-}$. The outer degree 16 rational map is constructed by taking zeros on the faces, poles on the edges, and both zeros and poles on the vertices. The points for the outer rational map are shown in Figure \ref{fig:B_20}, and the rational map is given by \cite{FLM}
\be
R_{20,T_d} = \left( \frac{1+c_2}{1+c_1} \right) \frac{p_+}{p_-} 
\left(\frac{c_1 p_+^3 + p_-^3}{p_+^3 + c_2 p_-^3}\right) \,,
\label{eqn:B_20map}
\ee
\noindent where $c_1=-2.873$ and $c_2=0.178$.
\clearpage

\begin{figure}[ht]
\centering
\includegraphics[width=0.3\textwidth]{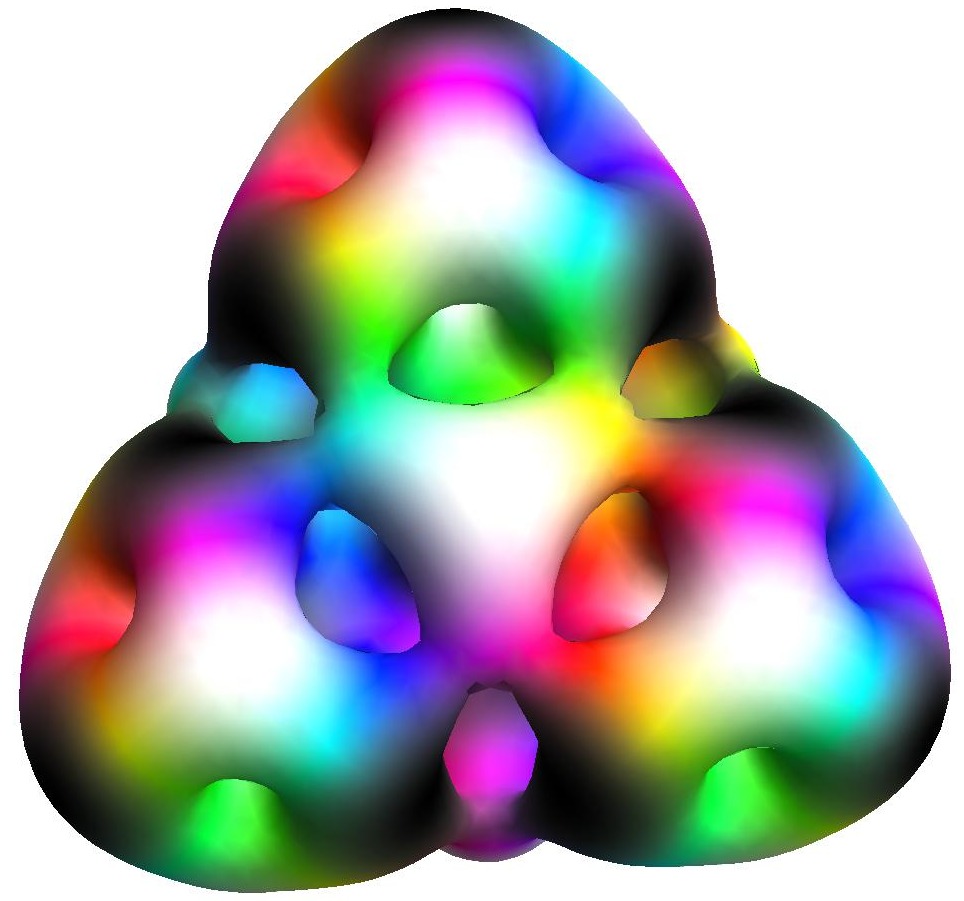}
\caption{$B=20$ $T_d$-symmetric Skyrmion}
\label{fig:B_20_Td}
\end{figure}

These rational maps give an initial ansatz. On relaxation we obtain the $B=20$ Skyrmion shown in Figure \ref{fig:B_20_Td} using the same colouring scheme as in \cite{FLM}.

\section{Quantization of $B=20$ $T_d$-symmetric Skyrmion}

In order to quantize the $B=20$ Skyrmion, we use the tensor product states $\left| L , L_3 \right \rangle \otimes \left| K, K_3 \right \rangle$ introduced in Section 2. The Finkelstein-Rubinstein (F-R) constraints are operator constraint equations which encode the symmetry and topology of the Skyrmion. For the $B=20$ $T_d$-symmetric Skyrmion, the F-R constraints are
\be
e^{\frac{2 \pi}{3 \sqrt{3}} (\hat{L}_1 + \hat{L}_2 + \hat{L}_3)} e^{i \frac{2\pi}{3} \hat{K}_3} \left | \Psi \right \rangle = \left | \Psi \right \rangle \, , \qquad \qquad e^{i \pi \hat{L}_3} \left | \Psi \right \rangle = \left | \Psi \right \rangle \, .
\label{eqn:FR}
\ee
\noindent The operators here are two generators of the tetrahedral rotation group, $T$. The reflection elements in $T_d$ will play a role when we consider the parity of states. The second constraint is simple; the eigenvalue of $\hat{L}_3$ must be even. The first constraint is more complicated and harder to solve. Two different ways of solving these constraints will be discussed in the following subsections.

Besides the isospin and spin, quantum states are labelled by parity. To construct the parity operator of the Skyrmion, the explicit form of the rational map $R(z)$ can be used. The parity operation on any Skyrmion is defined as a combined inversion $I$ in the complex coordinate, $\mathrm{z}$, and the isospace complex coordinate, $R$, i.e. a combination of $\mathrm{z} \rightarrow - \, \rfrac{1}{\bar{\mathrm{z}}}$ and $R \rightarrow - \, \rfrac{1}{\overline{R}}$.

Applying the combined inversion \textit{I} to the $B=20$ $T_d$-symmetric outer rational map (\ref{eqn:B_20map}), one finds
\be
R(\mathrm{z}) \xrightarrow{\quad \textit{I} \quad } -\frac{1}{\overline{R} (-\frac{1}{\bar{\mathrm{z}}})} = -\frac{1}{R(i\mathrm{z})} \, .
\ee
\noindent This shows that the parity operation on this Skyrmion is equivalent to a $\frac{\pi}{2}$-rotation about the $x_3$-axis ($\mathrm{z} \rightarrow i\mathrm{z}$) together with a $\pi$-rotation about the isospace 2-axis ($R \rightarrow - \, \rfrac{1}{R}$). The effect of $I$ on the inner rational map is the same. The quantum parity operator is therefore 
\be
\hat{P} = e^{i \frac{\pi}{2} \hat{L}_3} e^{i \pi \hat{K}_2} \,,
\label{eqn:par_op}
\ee
whose eigenvalue $\pm 1$ can be directly calculated for any state $\left| \Psi \right \rangle$. The parity operator can be understood more geometrically using the cubic grid, as will be shown in Subsection 4.4.

\subsection{Angular momentum basis method}

The allowed quantum states can be found by solving the F-R constraints, in an angular momentum basis. $L$ and $K$ are good quantum numbers, so we fix these. Dealing with the F-R constraints is a difficult task as it involves exponentiating matrices. With the help of \textit{Mathematica}, we found and solved the F-R constraints up to $L=4$, which requires exponentiating $9 \times 9$ matrices.

As an example, we apply this method to find the non-trivial quantum state with $L=3$ and $K=0$. For $K=0$, the F-R constraints simplify to
\be
e^{i \frac{2 \pi}{3 \sqrt{3}} (\hat{L}_1 + \hat{L}_2 + \hat{L}_3)} \left | \Psi \right \rangle = \left | \Psi \right \rangle \, , \qquad \qquad e^{i \pi \hat{L}_3} \left | \Psi \right \rangle = \left | \Psi \right \rangle \, .
\label{eqn:B20_FRK0}
\ee
\noindent We use the standard angular momentum basis, and write $\left | \Psi \right \rangle = \sum_{L_3=-L}^L C_{L_3} \left | L , L_3 \right \rangle$. The F-R constraints are
\begin{equation}
\begin{pmatrix}
    -\frac{i}{8} & - \frac{i}{4} \sqrt{\frac{3}{2}} & -\frac{i}{8} \sqrt{15} & -\frac{i}{4}\sqrt{5} &  -\frac{i}{8} \sqrt{15} & -  \frac{i}{4}\sqrt{\frac{3}{2}} & -\frac{i}{8} \\
    \frac{1}{4} \sqrt{\frac{3}{2}} & \frac{1}{2} & \frac{1}{4} \sqrt{\frac{5}{2}} & 0 & -\frac{1}{4} \sqrt{\frac{5}{2}} & -\frac{1}{2} & -\frac{1}{4} \sqrt{\frac{3}{2}} \\
    \frac{i}{8}\sqrt{15} & \frac{i}{4} \sqrt{\frac{5}{2}} & - \frac{i}{8} & -\frac{i}{4}\sqrt{3} &  - \frac{i}{8} & \frac{i}{4} \sqrt{\frac{5}{2}} & \frac{i}{8} \sqrt{15} \\
    -\frac{1}{4} \sqrt{5} & 0 & \frac{1}{4}\sqrt{3} & 0 & -\frac{1}{4}\sqrt{3} & 0 & \frac{1}{4}\sqrt{5} \\
    -\frac{i}{8}\sqrt{15} & \frac{i}{4} \sqrt{\frac{5}{2}} & \frac{i}{8} & -\frac{i}{4}\sqrt{3} &  \frac{i}{8} & \frac{i}{4} \sqrt{\frac{5}{2}} & -\frac{i}{8}\sqrt{15} \\
    \frac{1}{4} \sqrt{\frac{3}{2}} & -\frac{1}{2} & \frac{1}{4} \sqrt{\frac{5}{2}} & 0 & -\frac{1}{4} \sqrt{\frac{5}{2}} & \frac{1}{2} & -\frac{1}{4} \sqrt{\frac{3}{2}} \\
    \frac{i}{8} & - \frac{i}{4}\sqrt{\frac{3}{2}} & \frac{i}{8} \sqrt{15} & -\frac{i}{4}\sqrt{5} &  \frac{i}{8} \sqrt{15} & -  \frac{i}{4}\sqrt{\frac{3}{2}} & \frac{i}{8}
\end{pmatrix}
\begin{pmatrix}[1.4]
    C_{3} \\
    C_{2} \\
    C_{1} \\
    C_{0} \\
    C_{-1} \\
    C_{-2} \\
    C_{-3} \\
\end{pmatrix}
=
\begin{pmatrix}[1.4]
    C_{3} \\
    C_{2} \\
    C_{1} \\
    C_{0} \\
    C_{-1} \\
    C_{-2} \\
    C_{-3} \\
\end{pmatrix} \, ,
\end{equation}
\noindent and
\begin{equation}
\begin{pmatrix}
    -1 & 0 & 0 & 0 & 0 & 0 & 0 \\
    0 & 1 & 0 & 0 & 0 & 0 & 0 \\
    0 & 0 & -1 & 0 & 0 & 0 & 0 \\
    0 & 0 & 0 & 1 & 0 & 0 & 0 \\
    0 & 0 & 0 & 0 & -1 & 0 & 0 \\
    0 & 0 & 0 & 0 & 0 & 1 & 0 \\
    0 & 0 & 0 & 0 & 0 & 0 & -1 \\
\end{pmatrix}
\begin{pmatrix}
    C_{3} \\
    C_{2} \\
    C_{1} \\
    C_{0} \\
    C_{-1} \\
    C_{-2} \\
    C_{-3} 
\end{pmatrix}
=
\begin{pmatrix}
    C_{3} \\
    C_{2} \\
    C_{1} \\
    C_{0} \\
    C_{-1} \\
    C_{-2} \\
    C_{-3}
\end{pmatrix} \, .
\end{equation}
\indent The second can be solved without explicit calculation; only the states with even $L_3$ are allowed. These are $\left | 3 , 2 \right \rangle$, $\left | 3 , 0 \right \rangle$ and $\left | 3 , {-2} \right \rangle$ only. The first equation implies that there is only one allowed combination, namely
\be
\left| \Psi \right\rangle = \left( \left| 3 , 2 \right\rangle - \left| 3 , {-2} \right\rangle \right) \otimes \left| 0 , 0 \right\rangle \, ,
\ee  
\noindent where $\left | 0 , 0 \right \rangle$ is the isospin state. Applying the parity operator (\ref{eqn:par_op}), one finds that $\left| \Psi \right \rangle$ is a negative parity state, $3^-$. 

\subsection{Cartesian method}
The angular momentum basis method works well up to $L=4$ but becomes computationally intractable for higher spin. To overcome this, an alternative method using the Cartesian representation of spherical harmonics is developed. This method uses Cartesian coordinates ($x$,$y$,$z$) in ordinary space and ($X,Y,Z$) in isospace to form $T$-invariant polynomials. The Cartesian coordinates are set up with respect to the body-frames and are related to the Euler angles ($\theta,\psi$) and ($\beta,\gamma$) by
\begin{align}
x&=\rho\sin\theta \cos \psi & &,& \quad y&=\rho\sin\theta \sin \psi & &,& \quad z&=\rho\cos \theta &\, &; \\
X&=\lambda\sin\beta \cos \gamma & &,& \quad Y&=\lambda\sin\beta \sin \gamma & &,& \quad Z&=\lambda\cos \beta &\, &,
\end{align}
where $\rho^2=x^2+y^2+z^2$ and $\lambda^2=X^2+Y^2+Z^2$.

The F-R constraints (\ref{eqn:FR}) can be understood geometrically using the Cartesian coordinates. The operator $e^{i\pi \hat{L}_3}$ of the second constraint rotates ordinary space by $\pi$ about the $z$-axis (the ($0,0,1$)-direction) and there is no rotation in isospace. The Cartesian coordinates transform as
\begin{align}
x &\rightarrow -x & &, & \ y &\rightarrow -y & &, & \ z &\rightarrow z &\, &; \\
X &\rightarrow X & &, & Y &\rightarrow Y & &, & \ Z &\rightarrow Z &\, &.
\end{align}
The first constraint mixes both ordinary space and isospace. The operator $e^{i\frac{2\pi}{3\sqrt{3}}(\hat{L}_1+\hat{L}_2+\hat{L}_3)}$ rotates ordinary space by $\frac{2 \pi}{3}$ about the $(1,1,1)$-direction, and the operator $e^{i\frac{2\pi}{3}\hat{K}_3}$ rotates isospace by $\frac{2 \pi}{3}$ about the $Z$-axis. The effect on isospace is more transparent when we change variables from $X$ and $Y$ to $X+iY$ and $X-iY$. The Cartesian coordinates transform as
\begin{align}
x &\rightarrow y & &, & \ y &\rightarrow z & &, & \ z &\rightarrow x &\, &; \label{eqn:spin_perm} \\
 (X+iY) &\rightarrow \omega (X+iY) & &, & (X-iY) &\rightarrow \omega^2 (X-iY) & &, & \ Z &\rightarrow Z &\, &,
\end{align}
\noindent where $\omega = e^{i \frac{2 \pi}{3}}$.

The $3^-$ state with isospin zero has a simple Cartesian expression,
\begin{align}
\left| \Psi \right\rangle &= \left( \left| 3 , 2 \right\rangle - \left| 3 , {-2} \right\rangle \right) \otimes \left| 0 , 0 \right\rangle \, , \nonumber \\
&\propto (Y_{3,2}(\theta,\psi) - Y_{3,-2}(\theta,\psi))^* \otimes Y^*_{0,0}(\beta,\gamma) \, , \nonumber \\
&\propto \frac{1}{\rho^3} \left[ (x+iy)^2z - (x-iy)^2z \right] ^* \otimes 1 \, , \nonumber\\
&\propto \frac{1}{\rho^3} (x^2z+2ixyz- y^2z - x^2z+2ixyz + y^2z)^* \otimes 1 \, , \nonumber \\
&\propto \frac{1}{\rho^3} xyz \otimes 1 \, .
\end{align}
The state is proportional to a monomial of degree three, and is clearly invariant under (\ref{eqn:spin_perm}). $xyz$ is one of the $T$-invariant generating polynomials.
 
\subsubsection{Pure spin states}

We look here at the general F-R allowed states with isospin $K$ equal to zero. The $T$-invariant generating polynomials are listed in Table \ref{tab:spin}. These are further classified using $\hat{P}$, the parity operator (\ref{eqn:par_op}). The Cartesian coordinates transform under $\hat{P}$ as  
\begin{align}
\hat{P}: \quad x  &\rightarrow y & &,& \ y &\rightarrow {-x} & &,& \ z &\rightarrow z & \, &; \\
X  &\rightarrow -X & &,& \ Y &\rightarrow Y & &,& \ Z &\rightarrow -Z & \, &.
\end{align}
The first polynomial $f_2$ is $\rho^2$ and has zero spin. Dividing by $\rho^2$ gives $1$. The polynomial $f_3$ is the $L=3$ spin state that we presented, and $f_4$ is cubically-symmetric. The last polynomial $f_6$ is different from the others because it is not invariant under any reflection of $T_d$; however, it satisfies the F-R constraints and must be included.

\begin{table}[h]
\begin{center}
\begin{tabular}{ | c | c | c | c | }
\hline
Degree & & $T$-invariant polynomials & Parity \\ \hline
$2$ & $f_2$ & $x^2 + y^2 + z^2$ & $+$ \\ \hline
$3$ & $f_3$ & $xyz$ & $-$ \\ \hline
$4$ & $f_4$ & $x^4 + y^4 + z^4$ & $+$ \\ \hline
$6$ & $f_6$ & $x^4(y^2-z^2) + y^4(z^2-x^2) + z^4(x^2-y^2)$ & $-$ \\
\hline 
\end{tabular}
\end{center}
\caption{$T$-invariant polynomials of pure spin states}
\label{tab:spin}
\end{table}

To find a quantum state of spin $L$, a degree $L$ polynomial is constructed using the generating polynomials. As an example, an $L=4$ together with an $L=0$ state can be constructed using $f_2^2$ and $f_4$. The general degree 4 polynomial is
\be
F = a f_2^2 + b f_4 \, .
\label{eqn:F4}
\ee
With arbitrary constants $a$ and $b$, this is a state of mixed spin, but the ratio of the constants can be determined using 
\be
\hat{L}^2 \left | \Psi \right \rangle = L(L+1) \left | \Psi \right \rangle \, .
\label{eqn:L2}
\ee
The operator $\hat{L}^2$ is
\begin{align}
\hat{L}^2=& \, \hat{L}_x^2+\hat{L}_y^2+\hat{L}_z^2 \nonumber \\
=& - [ (y^2+z^2) \partial^2_x + (x^2+z^2) \partial^2_y + (x^2 + y^2) \partial^2_z \nonumber \\
& \ \ - 2(xy \partial_x \partial_y + xz \partial_x \partial_z + yz \partial_y \partial_z) -2(x \partial_x + y \partial_y + z \partial_z)] \, . \label{eqn:L2op}
\end{align}
where $(\hat{L}_x,\hat{L}_y,\hat{L}_z)$ are expressed in the Cartesian coordinates, for example, $\hat{L}_z= -i( x\partial_y-y \partial_x)$. For the polynomial (\ref{eqn:F4}), one obtains
\be
\hat{L}^2 F = 8 b x^4 - 24 b x^2 y^2 + 8 b y^4 - 24 b x^2 z^2 - 24 b y^2 z^2 + 8 b z^4 \,
\ee
and
\begin{align}
L(L+1)F = L(L+1)[&( a + b )x^4 + (a + b) y^4 + (a + b) z^4 \nonumber \\
& + 2 a x^2 y^2 + 2 a x^2 z^2 + 2 a y^2 z^2] \, ,
\end{align}
and to satisfy (\ref{eqn:L2}), two independent linear equations need to be satisfied, namely
\begin{align}
8b &= L(L+1) (a+b) \, , \\
-24b &= L(L+1) (2a) \, \label{eqn:L4K0} .
\end{align}
The solutions are $\frac{b}{a}=-\frac{5}{3}$ for $L=4$ and $b = 0$ for $L=0$. The pure spin states are
\be
F_{L=4,K=0}=3f_2^2 - 5f_4\, , \qquad F_{L=0,K=0}=f_2^2 \, .
\ee
\indent An alternative but equivalent way of fixing the constants of our polynomial quantum states is to use the Laplace equation, $\nabla^2f(x,y,z)=0$. The condition (\ref{eqn:L2}) is satisfied automatically by a degree $L$ polynomial $f$, if it is a solution to the Laplace equation. The proof is as follows. Let us consider a degree $L$ polynomial $f(x,y,z)= \sum_{i,j,k}a_{ijk} x^iy^jz^k$, where $a_{ijk}$ are constants, $i+j+k=L$ and $i,j,k\geq0$. In terms of the spherical polar coordinates, $f(x,y,z)$ is $\rho^Lg(\theta,\psi)$, where $g(\theta,\psi)$ is some angular function. Substituting this expression into the Laplace equation, expressed in spherical polars, one obtains
\begin{equation}
\nabla^2f =  \frac{1}{\rho^2}\left[\partial_\rho(\rho^2\partial_\rho f) - \hat{L}^2 f\right] = \ 0 \, .
\end{equation}
Because of the $\rho^L$ factor, $\partial_\rho(\rho^2\partial_\rho f) = L(L+1)f$, so $\nabla^2 f=0$ implies that
\begin{equation}
L(L+1)g - \hat{L}^2 g = 0 \, ,
\end{equation}
which means that $f$ is a state of spin $L$. The advantage of using the Laplace operator over (\ref{eqn:L2op}) is that it has a much simpler expression in terms of Cartesian coordinates,
\be
\nabla^2f=(\partial_x^2 + \partial_y^2 + \partial_z^2) f(x,y,z) \,,
\ee
and the calculation is easier.

\indent We can repeat the calculation for the $L=4$, $K=0$ state here using the Laplace operator. For $F=af_2^2+bf_4$,
\begin{align}
\nabla^2F&=\nabla^2\left[a(x^2+y^2+z^2)^2 + b(x^4+y^4+z^4) \right] \nonumber \\
&= (20a + 12b)(x^2 + y^2 + z^2) \nonumber \\
&= 0
\end{align}
if $\frac{b}{a}=-\frac{5}{3}$. The Laplace operator can only project out the quantum state with angular momentum equal to the degree of the polynomial; the other solution with $L=0$ cannot be obtained using this method. There is no $L=2$ state, because $f_2$, the only degree 2 polynomial, does not satisfy the Laplace equation.

States can be converted back to the angular momentum basis using the coefficients
\be
C_{L_3}= \iint F_{L,K}(\theta , \psi) Y^*_{L, L_3}(\theta,\psi) \sin \theta \, d \theta d \psi \, .
\label{eqn:ang_conv}
\ee
$F_{L,K}$ is expressed in spherical polar coordinates here and $Y_{L ,L_3}$ is the standard spherical harmonic. The spin 4 state $F=3f_2^2 - 5f_4$ in spherical polar coordinates is
\be
F_{L=4,K=0}= 3 - 5 \Big( \sin^4 \theta \cos ^4 \psi + \sin^4 \theta \sin ^4 \psi + \cos ^4 \theta \Big) \, .
\ee
Using equation (\ref{eqn:ang_conv}), the state can be expressed in the angular momentum basis as
\begin{align}
\left | \Psi \right \rangle &= \sum_{L_3=-L}^L C_{L_3} \left | L , L_3 \right \rangle \otimes \left| 0,0 \right \rangle \nonumber \\
&= \left(\left | 4 , -4 \right \rangle + \sqrt{\frac{14}{5}} \left | 4, 0 \right \rangle + \left | 4 , 4 \right \rangle \right) \otimes \left | 0, 0 \right \rangle \, ,
\end{align}
which involves an irrational coefficient. This is to be compared with the neat rational coefficients of the Cartesian form of the state.
\subsubsection{Pure isospin states}

We now move to the case of pure isospin with spin $L$ set to zero. The approach is very similar, and the invariant generating polynomials are presented in Table \ref{tab:isospin}. 

\begin{table}[!h]
\begin{center}
\begin{tabular}{ | c | c | c | c | }
\hline
Degree & & & Parity  \\ \hline
$1$ & $g_1$ & $Z$ & $-$  \\ \hline
$2$ & $g_2$ & $X^2 + Y^2$ & $+$  \\ \hline
$3$ & $g_{3-}$ & $X(X^2-3Y^2)$ & $-$ \\ \hline
$3$ & $g_{3+}$ & $Y(3X^2-Y^2)$ & $+$ \\
\hline 
\end{tabular}
\end{center}
\caption{Invariant generating polynomials of pure isospin states}
\label{tab:isospin}
\end{table}

As an example, we construct the $K=3$ states using degree 3 polynomials. For degree 3, four polynomials are allowed, ($g_1^3$, $g_2g_1$, $g_{3-}$ and $g_{3+}$), compared to only two, ($f_2^2,f_4$), for degree 4 in the case of pure spin. The calculation can be simplified with the help of parity. The only positive parity polynomial is $g_{3+}$, and it is the only possible positive parity $K=3$ state. The negative parity polynomials are $g_1^3$, $g_2g_1$ and $g_{3-}$. To find pure isospin states, the same procedure as in the case of pure spin is repeated: one formulates and solves simultaneous equations by comparing the polynomials term by term. The $K=3$ states are
\begin{align}
F^+_{0,3}&=g_{3+} \, , \\
F^-_{0,3}&=g_1^3-\frac{3}{2}g_2g_1 \, .
\end{align}

\subsubsection{Mixed spin and isospin}

The cases of pure spin and pure isospin have been covered. We wish to construct mixed quantum states too. Here we present all states up to spin 4 and isospin 2. $L_3$ must be even, as before. The F-R constraint
\be
e^{i \frac{2 \pi}{3 \sqrt{3}} (\hat{L}_1 + \hat{L}_2 + \hat{L}_3)} e^{i \frac{2\pi}{3} \hat{K}_3} \left | \Psi \right \rangle = \left | \Psi \right \rangle \,
\label{eqn:FReqn_2}
\ee
\noindent allows the spin and isospin operators each to have a complex eigenvalue of unit magnitude. The constraint equation can be satisfied as long as the phases cancel out. This allows two more generating polynomials for both spin and isospin. They are
\be
\omega(x^2+\omega y^2 +\omega^2 z^2), \ \omega^2(x^2 + \omega^2 y^2 + \omega z^2), \ (X+iY) \text{ and } \ (X-iY) \, ,
\ee
\noindent where $\omega = e^{i \frac{2\pi}{3}}$ as before. Notice that they are not eigenstates of the parity operator but transform into each other under parity (see Table \ref{tab:s_and_iso}).

\begin{table}[ht]
\begin{center}
\begin{tabular}{ | c | c | c | c | c | }
\hline
Degree & & & Under F-R (\ref{eqn:FReqn_2}) & Under parity \\ \hline
$2$ & $f_{\omega}$ & $\omega^2(x^2 + \omega^2 y^2 + \omega z^2)$ & $\omega \, f_{\omega}$ & $f_{\omega^2}$ \\ \hline
$2$ & $f_{\omega^2}$ & $\omega(x^2+\omega y^2 +\omega^2 z^2)$ & $\omega^2 f_{\omega^2}$ & $f_{\omega}$ \\ \hline
$1$ & $g_{\omega}$ & $(X+iY)$ & $\omega \, g_{\omega}$ & $-g_{\omega^2}$ \\ \hline
$1$ & $g_{\omega^2}$ & $(X-iY)$ & $\omega^2 g_{\omega^2}$ & $-g_{\omega}$ \\
\hline 
\end{tabular}
\end{center}
\caption{Generating polynomials for mixed spin and isospin states}
\label{tab:s_and_iso}
\end{table}

We can now start constructing quantum states satisfying the F-R constraints with both non-zero spin and isospin; the allowed terms are tensor products of the spin and isospin generating polynomials where $\omega$ and $\omega^2$ factors cancel. $f_\omega$ and $f_{\omega^2}$ satisfy the Laplace equation, hence they are states of pure spin with $L=2$. As an example, the allowed terms for $L=2,K=1$ states are
\be
f_{\omega} \otimes g_{\omega^2} \text{ and } f_{\omega^2} \otimes g_{\omega} \, .
\ee
\noindent Note that the zero spin polynomial $f_2$ is not needed. We obtain the following two states with definite parities
\begin{align}
F^{+}_{L=2,K=1} &= f_{\omega} \otimes g_{\omega^2} - f_{\omega^2} \otimes g_{\omega} \, ; \\
F^{-}_{L=2,K=1} &= f_{\omega} \otimes g_{\omega^2} + f_{\omega^2} \otimes g_{\omega} \, .
\end{align}
The list of the allowed quantum states for the $B=20$ $T_d$-symmetric Skyrmion up to $L=4$, $K=2$ is presented in Table \ref{tab:quantum}. The states are arranged according to the spin and do not reflect the actual ordering of energy levels. In order to evaluate the energy of each state, the moments of inertia are required. The energy calculation and the list of energy levels are discussed in the following subsection.

\begin{table}[ht]
\begin{center}
\hspace*{-0.8cm}
\begin{tabular}{ | c | c | p{8.4cm} | p{4cm} | }
\hline
Spin, $L^{P}$ & Isospin, $K$ & $\left| \Psi \right \rangle$ in angular momentum basis & $\left| \Psi \right \rangle$ in Cartesian basis\\ \hline
$0^{+}$ & $0$ & $\left | 0, 0 \right \rangle \otimes \left | 0, 0 \right \rangle$ & $1$ \\ \hline
$3^{-}$ & $0$ & $\left | 3, \pm 2 \right \rangle_{-} \otimes \left | 0, 0 \right \rangle $ & $f_3$ \\ \hline
$4^{+}$ & $0$ & $\left( \left | 4 ,\pm 4 \right \rangle_{+} + \sqrt{\frac{14}{5}} \left | 4, 0 \right \rangle \right) \otimes \left | 0, 0 \right \rangle$ & $f_2^2-\frac{5}{3}f_4$ \\ \hline
$0^{-}$ & $1$ & $\left | 0, 0 \right \rangle \otimes \left | 1 , 0 \right \rangle $ & $g_1$ \\ \hline
$2^{+}$ & $1$ & $\left( \left | 2 , \pm 2 \right \rangle_{+} + \sqrt{2}i \left | 2 , 0 \right \rangle \right) \otimes \left | 1 , -1 \right \rangle - \newline \left( \left | 2 , \pm 2 \right \rangle_{+} - \sqrt{2}i \left | 2 , 0 \right \rangle \right) \otimes \left | 1 , 1 \right \rangle$ & $f_{\omega}g_{\omega^2}-f_{\omega^2}g_{\omega}$ \\ \hline
$2^{-}$ & $1$ & $\left( \left | 2 , \pm 2 \right \rangle_{+} + \sqrt{2}i \left | 2 , 0 \right \rangle \right) \otimes \left | 1 , -1 \right \rangle + \newline \left( \left | 2 , \pm 2 \right \rangle_{+} - \sqrt{2}i \left | 2 , 0 \right \rangle \right) \otimes \left | 1 , 1 \right \rangle$ & $f_{\omega}g_{\omega^2}+f_{\omega^2}g_{\omega}$ \\ \hline
$3^{+}$ & $1$ & $\left | 3 , \pm 2 \right \rangle_{-}  \otimes \left | 1 , 0 \right \rangle $ & $f_{3}g_{1}$ \\ \hline
$4^{+}$ & $1$ & $\left( \left | 4 , \pm 4 \right \rangle_{+} + \sqrt{\frac{12}{7}}i \left | 4 , \pm 2 \right \rangle_{+} - \frac{10}{7} \left | 4 , 0 \right \rangle \right) \otimes \left | 1 , 1 \right \rangle \newline  + \newline \left( \left | 4 , \pm 4 \right \rangle_{+} - \sqrt{\frac{12}{7}}i \left | 4 , \pm 2 \right \rangle_{+} - \frac{10}{7} \left | 4 , 0 \right \rangle \right) \otimes \left | 1 , -1 \right \rangle$ & $f_2(f_{\omega}g_{\omega^2} - f_{\omega^2}g_{\omega}) + \frac{7}{4} (f_{\omega}^2 g_{\omega} - f_{\omega^2}^2 g_{\omega^2})$ \\ \hline
$4^{-}$ & $1$ & $\left( \left | 4 , \pm 4 \right \rangle_{+} + \sqrt{\frac{12}{7}}i \left | 4 , \pm 2 \right \rangle_{+} - \frac{10}{7} \left | 4 , 0 \right \rangle \right) \otimes \left | 1 , 1 \right \rangle \newline - \newline \left( \left | 4 , \pm 4 \right \rangle_{+} - \sqrt{\frac{12}{7}}i \left | 4 , \pm 2 \right \rangle_{+} - \frac{10}{7} \left | 4 , 0 \right \rangle \right) \otimes \left | 1 , -1 \right \rangle$ & $f_2(f_{\omega}g_{\omega^2} + f_{\omega^2}g_{\omega}) - \frac{7}{4} (f_{\omega}^2 g_{\omega} + f_{\omega^2}^2 g_{\omega^2})$ \\ \hline
$4^{-}$ & $1$ & $\left( \left | 4 ,\pm 4 \right \rangle_{+} + \sqrt{\frac{14}{5}} \left | 4, 0 \right \rangle \right) \otimes \left | 1, 0 \right \rangle$ & $f_2^2g_1-\frac{5}{3}f_4 g_1$ \\ \hline
$0^{+}$ & $2$ & $\left | 0, 0 \right \rangle \otimes \left | 2 , 0 \right \rangle $ & $g_1^2 - \frac{1}{2}g_2$ \\ \hline
$2^{+}$ & $2$ & $\left( \left | 2 , \pm 2 \right \rangle_{+} + \sqrt{2}i \left | 2 , 0 \right \rangle \right) \otimes \left | 2 , -1 \right \rangle + \newline \left( \left | 2 , \pm 2 \right \rangle_{+} - \sqrt{2}i \left | 2 , 0 \right \rangle \right) \otimes \left | 2 , 1 \right \rangle$ & $f_{\omega}g_{\omega^2}g_1 + f_{\omega^2}g_{\omega}g_1$ \\ \hline
$2^{+}$ & $2$ & $\left( \left | 2 , \pm 2 \right \rangle_{+} + \sqrt{2}i \left | 2 , 0 \right \rangle \right) \otimes \left | 2 , 2 \right \rangle - \newline \left( \left | 2 , \pm 2 \right \rangle_{+} - \sqrt{2}i \left | 2 , 0 \right \rangle \right) \otimes \left | 2 , -2 \right \rangle$ & $f_{\omega}g_{\omega}^2 + f_{\omega^2}g_{\omega^2}^2$ \\ \hline
$2^{-}$ & $2$ & $\left( \left | 2 , \pm 2 \right \rangle_{+} + \sqrt{2}i \left | 2 , 0 \right \rangle \right) \otimes \left | 2 , -1 \right \rangle - \newline \left( \left | 2 , \pm 2 \right \rangle_{+} - \sqrt{2}i \left | 2 , 0 \right \rangle \right) \otimes \left | 2 , 1 \right \rangle$ & $f_{\omega}g_{\omega^2}g_1 - f_{\omega^2}g_{\omega}g_1$ \\ \hline
$2^{-}$ & $2$ & $\left( \left | 2 , \pm 2 \right \rangle_{+} + \sqrt{2}i \left | 2 , 0 \right \rangle \right) \otimes \left | 2 , 2 \right \rangle + \newline \left( \left | 2 , \pm 2 \right \rangle_{+} - \sqrt{2}i \left | 2 , 0 \right \rangle \right) \otimes \left | 2 , -2 \right \rangle$ & $f_{\omega}g_{\omega}^2 - f_{\omega^2}g_{\omega^2}^2$ \\ \hline
$3^{-}$ & $2$ & $\left | 3 , \pm 2 \right \rangle_{-}  \otimes \left | 2 , 0 \right \rangle $ & $f_{3}(g_{1}^2-\frac{1}{2}g_2)$ \\ \hline
$4^{+}$ & $2$ & $\left( \left | 4 ,\pm 4 \right \rangle_{+} + \sqrt{\frac{14}{5}} \left | 4, 0 \right \rangle \right) \otimes \left | 2, 0 \right \rangle$ & $(f_2^2-\frac{5}{3}f_4)(g_1^2-\frac{1}{2}g_2)$ \\ \hline
$4^{+}$ & $2$ & $\left( \left | 4 , \pm 4 \right \rangle_{+} + \sqrt{\frac{12}{7}}i \left | 4 , \pm 2 \right \rangle_{+} - \frac{10}{7} \left | 4 , 0 \right \rangle \right) \otimes \left | 2 , -2 \right \rangle \newline - \newline \left( \left | 4 , \pm 4 \right \rangle_{+} - \sqrt{\frac{12}{7}}i \left | 4 , \pm 2 \right \rangle_{+} - \frac{10}{7} \left | 4 , 0 \right \rangle \right) \otimes \left | 2 , 2 \right \rangle$ & $f_2(f_{\omega}g_{\omega}^2 + f_{\omega^2}g_{\omega^2}^2) - \frac{7}{4} (f_{\omega}^2 g_{\omega^2}^2 + f_{\omega^2}^2 g_{\omega}^2)$ \\ 
\hline
$4^{-}$ & $2$ & $\left( \left | 4 , \pm 4 \right \rangle_{+} + \sqrt{\frac{12}{7}}i \left | 4 , \pm 2 \right \rangle_{+} - \frac{10}{7} \left | 4 , 0 \right \rangle \right) \otimes \left | 2 , -2 \right \rangle \newline  + \newline \left( \left | 4 , \pm 4 \right \rangle_{+} - \sqrt{\frac{12}{7}}i \left | 4 , \pm 2 \right \rangle_{+} - \frac{10}{7} \left | 4 , 0 \right \rangle \right) \otimes \left | 2 , 2 \right \rangle$ & $f_2(f_{\omega}g_{\omega}^2 - f_{\omega^2}g_{\omega^2}^2) + \frac{7}{4} (f_{\omega}^2 g_{\omega^2}^2 - f_{\omega^2}^2 g_{\omega}^2)$ \\
\hline 
\end{tabular}
\end{center}
\caption{Allowed quantum states $\left| \Psi \right \rangle$ of the $B=20$ $T_d$-symmetric Skyrmion. The notation used is $\left | L , \pm L_3 \right \rangle_{\pm} = \left | L , L_3 \right \rangle \pm \left | L , -L_3 \right \rangle$.}
\label{tab:quantum}
\end{table}

\clearpage
\subsection{Energy calculation}

The energy levels of the $B=20$ $T_d$-symmetric Skyrmion can be computed using the moments of inertia given in Table \ref{tab:B_20} in the Appendix. Recall that the Hamiltonian of the Skyrmion is
\be
\hat{H}=\frac{1}{2v}\hat{L}^2 + \frac{1}{2U_{11}}(\hat{K}^2-\hat{K}_3^2) +\frac{1}{2U_{33}}\hat{K}_3^2 \, .
\label{eqn:20Ham}
\ee
From Table \ref{tab:B_20}, we extract the values $v=12854$, $U_{11}=759$ and $U_{33}=820$.

We illustrate the energy calculation using the $3^-$ state, $ \left| \Psi \right \rangle = \left| 3, \pm 2 \right \rangle _{-} \otimes \left| 0, 0 \right \rangle$. It is an eigenstate of the operators, $\hat{L}^2$, $\hat{L}_3^2$, $\hat{K}^2$ and $\hat{K}_3^2$, hence an eigenstate of the Hamiltonian. The energy $E$, in Skyrme units, can be calculated by replacing the operators with their eigenvalues,
\begin{align}
E_{3,0}&= \frac{1}{2v}L(L+1) + \frac{1}{2U_{11}}(K(K+1) - K_3^2) + \frac{1}{2U_{33}} K_3^2 \nonumber \\
&= \frac{3(3+1)}{2v} + 0 + 0 \nonumber \\
&= \frac{6}{v} \nonumber \\
&= 4.6677 \times 10^{-4} \,.
\end{align}

Using the conversion factor $e^3F_\pi=4000$ MeV, the state has an energy of $1.87$ MeV above the ground state. This is the correct order of magnitude for the excitation energy of a nucleus of this size.

A list of energy levels of the $B=20$ $T_d$-symmetric Skyrmion up to spin $4$ and isospin $2$ is presented in Table \ref{tab:energy}.
\begin{table}[ht]
\begin{center}
\begin{tabular}{ | c | c | c | c | c | }
\hline
Spin, $L^{P}$ & Isospin, $K$ & Energy, $E$ & $E$ in Skyrme units & $E$ in MeV \\ \hline
$0^{+}$ & $0$ & $0$ & $0$ & $0$ \\ \hline
$3^{-}$ & $0$ & $\frac{6}{v}$ & $4.668 \times 10^{-4}$ & $1.87$ \\ \hline
$4^{+}$ & $0$ & $\frac{10}{v}$ & $7.780 \times 10^{-4}$ & $3.11$ \\ \hline
$0^{-}$ & $1$ & $\frac{1}{U_{11}}$ & $13.184 \times 10^{-4}$ & $5.27$ \\ \hline
$2^{+}$ & $1$ & $\frac{3}{v}+\frac{1}{2U_{11}}+\frac{1}{2U_{33}}$ & $15.024 \times 10^{-4}$ & $6.01$ \\ \hline
$2^{-}$ & $1$ & $\frac{3}{v}+\frac{1}{2U_{11}}+\frac{1}{2U_{33}}$ & $15.024 \times 10^{-4}$ & $6.01$ \\ \hline
$3^{+}$ & $1$ & $\frac{6}{v}+\frac{1}{U_{11}}$ & $17.851 \times 10^{-4}$ & $7.14$ \\ \hline
$4^{+}$ & $1$ & $\frac{10}{v}+\frac{1}{2U_{11}}+\frac{1}{2U_{33}}$ & $20.470\times 10^{-4}$ & $8.19$ \\ \hline
$4^{-}$ & $1$ & $\frac{10}{v}+\frac{1}{2U_{11}}+\frac{1}{2U_{33}}$ & $20.470\times 10^{-4}$ & $8.19$ \\ \hline
$4^{-}$ & $1$ & $\frac{10}{v}+\frac{1}{U_{11}}$ & $20.963\times 10^{-4}$ & $8.39$ \\ \hline
$0^{+}$ & $2$ & $\frac{3}{U_{11}}$ & $39.551\times 10^{-4}$ & $15.82$ \\ \hline
$2^{+}$ & $2$ & $\frac{3}{v}+\frac{1}{U_{11}}+\frac{2}{U_{33}}$ & $39.910\times 10^{-4}$ & $15.96$ \\ \hline
$2^{-}$ & $2$ & $\frac{3}{v}+\frac{1}{U_{11}}+\frac{2}{U_{33}}$ & $39.910\times 10^{-4}$ & $15.96$ \\ \hline
$2^{+}$ & $2$ & $\frac{3}{v}+\frac{5}{2U_{11}}+\frac{1}{2U_{33}}$ & $41.391\times 10^{-4}$ & $16.56$ \\ \hline
$2^{-}$ & $2$ & $\frac{3}{v}+\frac{5}{2U_{11}}+\frac{1}{2U_{33}}$ & $41.391\times 10^{-4}$ & $16.56$ \\ \hline
$3^{-}$ & $2$ & $\frac{6}{v}+\frac{3}{U_{11}}$ & $44.219\times 10^{-4}$ & $17.69$ \\ \hline
$4^{+}$ & $2$ & $\frac{10}{v}+\frac{1}{U_{11}}+\frac{2}{U_{33}}$ & $45.356\times 10^{-4}$ & $18.14$ \\ 
\hline
$4^{-}$ & $2$ & $\frac{10}{v}+\frac{1}{U_{11}}+\frac{2}{U_{33}}$ & $45.356\times 10^{-4}$ & $18.14$ \\
\hline
$4^{+}$ & $2$ & $\frac{10}{v} + \frac{3}{U_{11}}$ & $47.331\times 10^{-4}$ & $18.93$ \\ \hline
\end{tabular}
\end{center}
\caption{Energy levels of $B=20$ $T_d$-symmetric Skyrmion}
\label{tab:energy}
\end{table}

\subsection{Construction of parity operator}

In this subsection, a pictorial method of constructing the parity operator $\hat{P}$ is presented. The effect of combined inversions in both ordinary space and isospace on a rational map,
\be
R(\mathrm{z}) \rightarrow - \frac{1}{\overline{R}(-\frac{1}{\bar{\mathrm{z}}})} \, ,
\ee
is first studied algebraically, using a simple example, then we demonstrate that it can be equivalently represented pictorially, using the cubic grid, and the corresponding parity operator can be read off from the grid without any algebraic calculation.

The simple example is a rational map constructed from two pairs of zero/pole points, ($a,- \, \rfrac{1}{\bar{a}}$) and ($\rfrac{1}{a},-\bar{a}$). The two pairs are related by a $\pi$-rotation about the $x_1$-axis ($z \rightarrow \rfrac{1}{z}$), and the points of each pair are antipodal. There is no tetrahedral symmetry here. The rational map is 
\be
R(\mathrm{z})=\frac{(\mathrm{z}-a)(\mathrm{z}-\frac{1}{a})}{(\mathrm{z}-(-\frac{1}{\bar{a}}))(\mathrm{z}-(-\bar{a}))} \, ,
\label{eqn:sim_rat}
\ee
and is transformed under the combined inversion $I$ to
\begin{align}
R(\mathrm{z}) \xrightarrow{\quad \textit{\large{$I$}} \quad} -\frac{1}{\overline{R}(-\frac{1}{\bar{\mathrm{z}}})} &= -\frac{\overline{(-\frac{1}{\bar{\mathrm{z}}}-(-\frac{1}{\bar{a}}))(-\frac{1}{\bar{\mathrm{z}}}-(-\bar{a}))}}{\overline{(-\frac{1}{\bar{\mathrm{z}}}-a)(-\frac{1}{\bar{\mathrm{z}}}-\frac{1}{a})}} \nonumber \\
&= -\frac{(-a\bar{a})(\mathrm{z}-a)(\mathrm{z}-\frac{1}{a})}{(-a\bar{a})(\mathrm{z}-(-\frac{1}{\bar{a}}))(\mathrm{z}-(-\bar{a}))} \nonumber \\
&= -R(\mathrm{z}) \, .
\label{eqn:sim_alg}
\end{align}
$R(\mathrm{z})$ is almost invariant, but gains a minus sign, so $I$ is equivalent here to a $\pi$-rotation about the $Z$-axis. The parity operator is $\hat{P}=e^{i \pi \hat{K}_3}$ in this case.

\begin{figure}[ht]
\noindent\begin{minipage}{.25\textwidth}
  \centering
  \includegraphics[width=\textwidth]{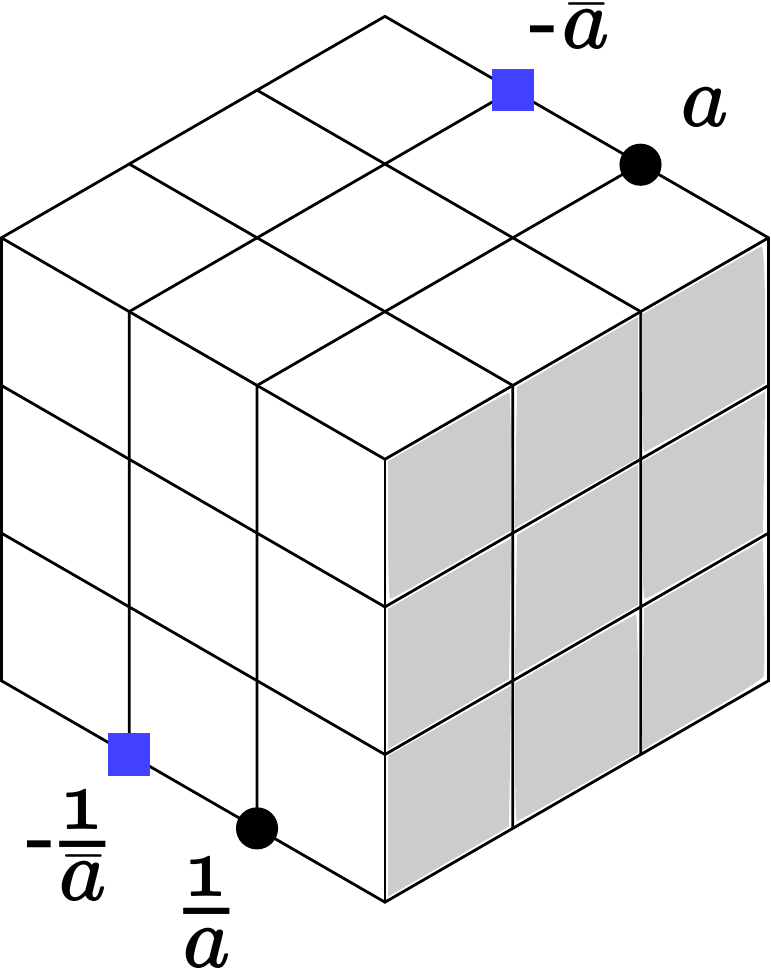}
\end{minipage}
\begin{minipage}{.1\textwidth}
\be
\xrightarrow{\; \text{ \; \large{$I_{ \textrm{s} }$ } \;}} \nonumber
\ee
\end{minipage}
\begin{minipage}{.25\textwidth}
  \centering
  \includegraphics[width=\textwidth]{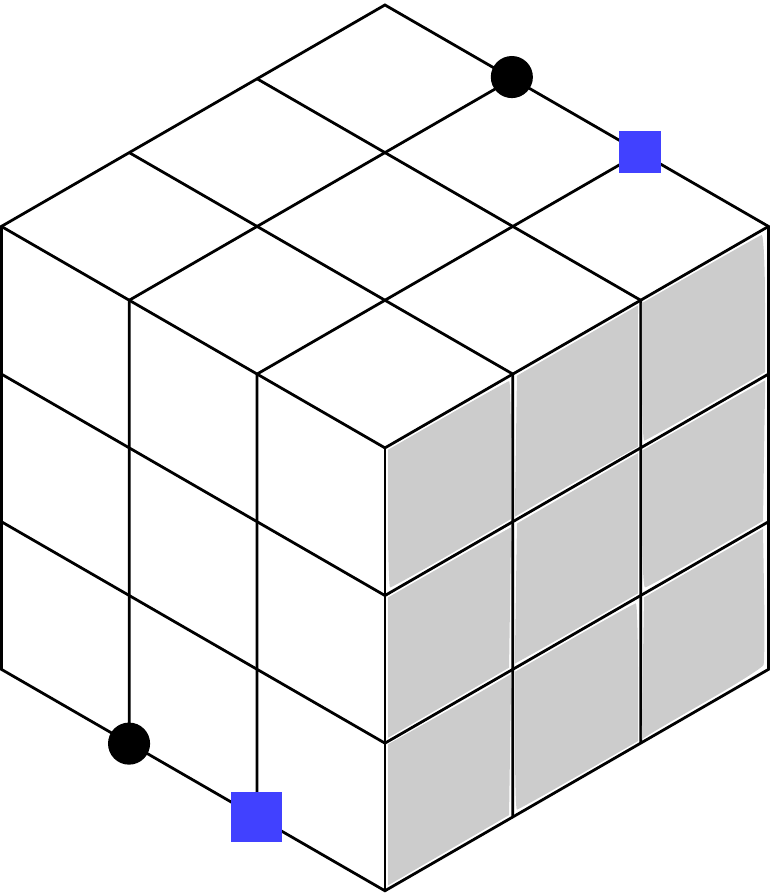}
\end{minipage}
\begin{minipage}{.1\textwidth}
\be
\xrightarrow{\text{ \large{$\sigma_{ \textrm{iso} }$} }} \nonumber
\ee
\end{minipage}
\begin{minipage}{.25\textwidth}
  \centering
  \includegraphics[width=\textwidth]{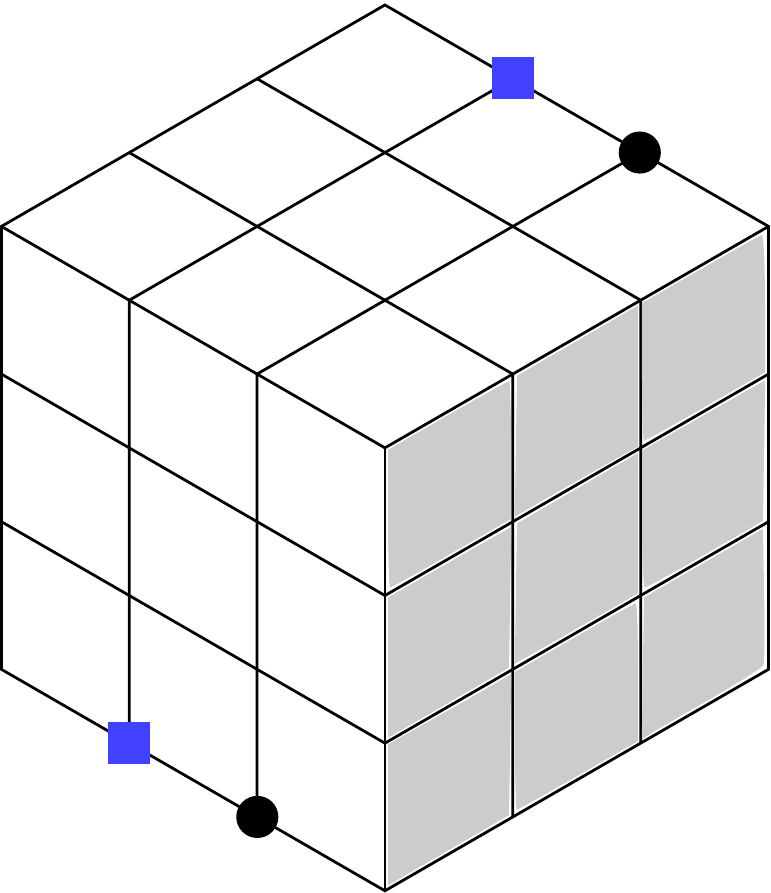}
\end{minipage}
\caption{Effect on the grid under $I_{ \textrm{s} }$ followed by $\sigma_{ \textrm{iso} }$}
\label{fig:parity_example}
\end{figure}

To explain our pictorial method, the points above are, for illustrative purposes only, assumed to be on the edges of the grid. The effect of the combined inversion on the rational map (\ref{eqn:sim_rat}) is represented in Figure \ref{fig:parity_example}. A spatial inversion $I_{\textrm{s}}$ ($\mathrm{z} \rightarrow - \, \rfrac{1}{\bar{\mathrm{z}}}$) is applied first; it moves the points to the antipodal points and exchanges zeros and poles. Second, an isospatial inversion is applied; this is a combined reflection $\sigma_{\textrm{iso}}$ in the $X-Y$ plane with a $\pi$-rotation, $C_{2 \, \textrm{iso}}$, about the $Z$-axis. The reflection $\sigma_{\textrm{iso}}$ exchanges zeros and poles ($R \rightarrow \rfrac{1}{\overline{R}}$); the isorotation $C_{2 \, \textrm{iso}}$ multiplies the rational map by $-1$ ($R \rightarrow -R$), but it does not change the positions of zeros and poles; as a result, the operation $C_{2 \, \textrm{iso}}$ cannot be represented in the Figure.

From Figure \ref{fig:parity_example}, one sees that the rational map returns to itself exactly after the operations $I_{\textrm{s}}$ and $\sigma_{\textrm{iso}}$. Including the factor $-1$ from the isorotation $C_{2 \, \textrm{iso}}$, we arrive at the same expression as shown in the algebraic calculation (\ref{eqn:sim_alg}),
\be
R(\mathrm{z}) \xrightarrow{\quad \text{\large{$I_{ \textrm{s}}$ , $\sigma_{\textrm{iso}}$ }} \quad} \frac{1}{\overline{R}(-\frac{1}{\bar{\mathrm{z}}})} = R(\mathrm{z}) \xrightarrow{\quad \text{\large{$C_{2 \, \textrm{iso}}$}} \quad} -R(\mathrm{z}) \, ,
\ee
and the parity operator $\hat{P}=e^{i \pi \hat{K}_3}$ is recovered.

\begin{figure}[ht]
\noindent\begin{minipage}{.25\textwidth}
  \centering
  \includegraphics[width=\textwidth]{4x4x4_Grid_Td.pdf}
  \label{fig:Td_grid_1}
\end{minipage}
\begin{minipage}{.1\textwidth}
\be
\xrightarrow{\; \text{  \large{$I_{\textrm{s}}$}  } \;} \nonumber
\ee
\end{minipage}
\begin{minipage}{.25\textwidth}
  \centering
  \includegraphics[width=\textwidth]{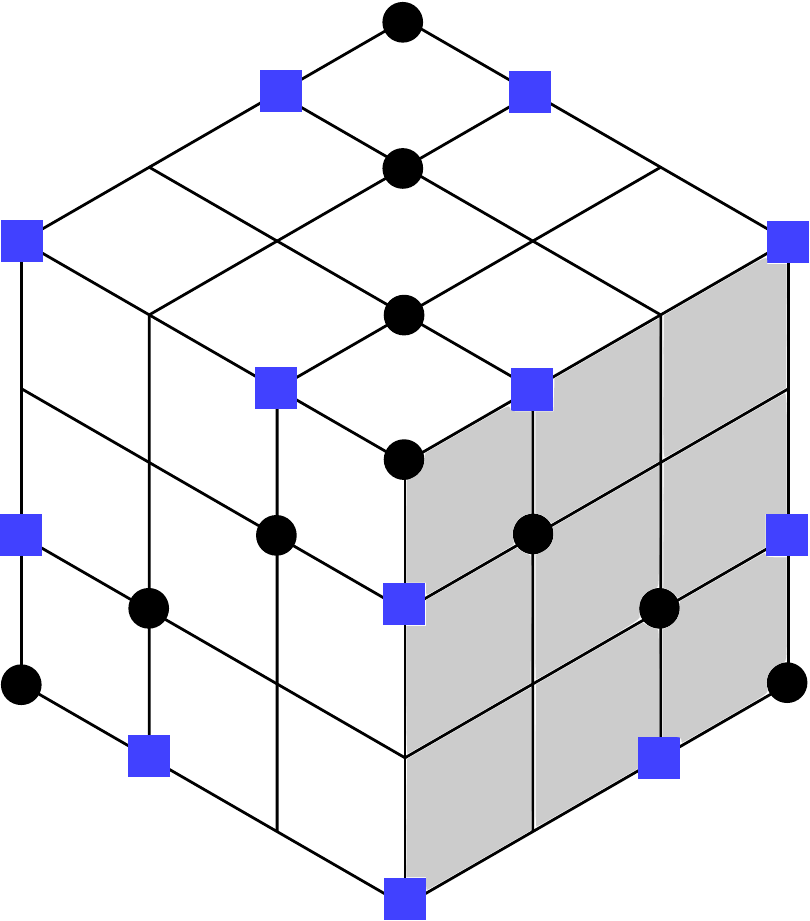}
  \label{fig:Td_grid_2}
\end{minipage}
\begin{minipage}{.1\textwidth}
\be
\xrightarrow{\text{ \large{$\sigma_{\textrm{iso}}$} }} \nonumber
\ee
\end{minipage}
\begin{minipage}{.25\textwidth}
  \centering
  \includegraphics[width=\textwidth]{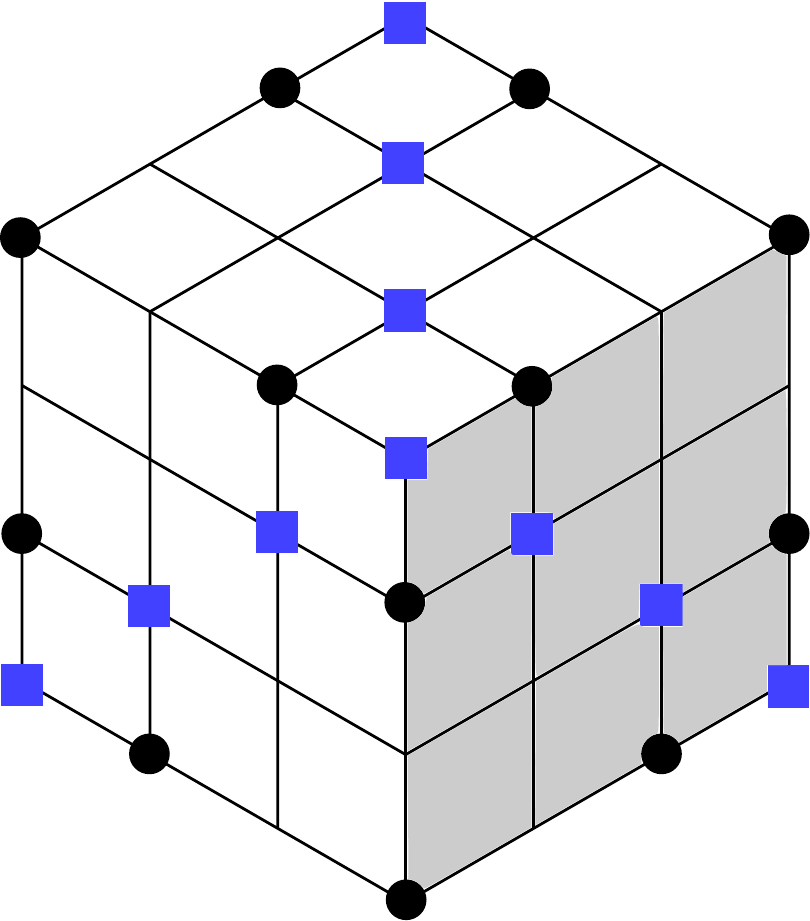}
  \label{fig:Td_grid_3}
\end{minipage}
\caption{$B=20$ rational map transformed by $I_{\textrm{s}}$ followed by $\sigma_{\textrm{iso}}$}
\label{fig:B20parity}
\end{figure}

The algebraic calculation becomes messy for rational maps of high degree. The pictorial method is then particularly effective. We now apply this method to the rational map for the $B=20$ $T_d$-symmetric Skyrmion; the operations are shown in Figure \ref{fig:B20parity}.

The rational map does not return to its original form after $I_{\textrm{s}}$ and $\sigma_{\textrm{iso}}$; the transformed map is related to the original map by a $\frac{\pi}{2}$-rotation about the $x_3$-axis ($\textrm{z} \rightarrow i\textrm{z}$) and a $\pi$-rotation about the $X$-axis ($R \rightarrow \rfrac{1}{R}$). We deduce that, including the isorotation $C_{2 \, \textrm{iso}}$, the inversions have the following algebraic effect
\be
R(\textrm{z}) \xrightarrow{\quad \text{\large{$I_{ \textrm{s}}$ , $\sigma_{\textrm{iso}}$ }} \quad} \frac{1}{\overline{R}(-\frac{1}{\bar{\textrm{z}}})} = \frac{1}{R(i\textrm{z})} \xrightarrow{\quad \text{\large{$C_{2 \, \textrm{iso}}$}} \quad} -\frac{1}{R(i\textrm{z})} \, ,
\ee
and the parity operator $\hat{P}=e^{i \frac{\pi}{2} \hat{L}_3}e^{i \pi \hat{K}_2}$ can be read off. ($C_{2 \, \textrm{iso}}$ has the effect of replacing the $\pi$-rotation about $X$-axis by a $\pi$-rotation about the $Y$-axis.)

\section{Further $T_d$-symmetric Skyrmions}

\subsection{Quantization, F-R constraints and parity operator}
\begin{figure}[ht]
\centering
\noindent\begin{minipage}{.3\textwidth}
  \centering
  \includegraphics[width=\textwidth]{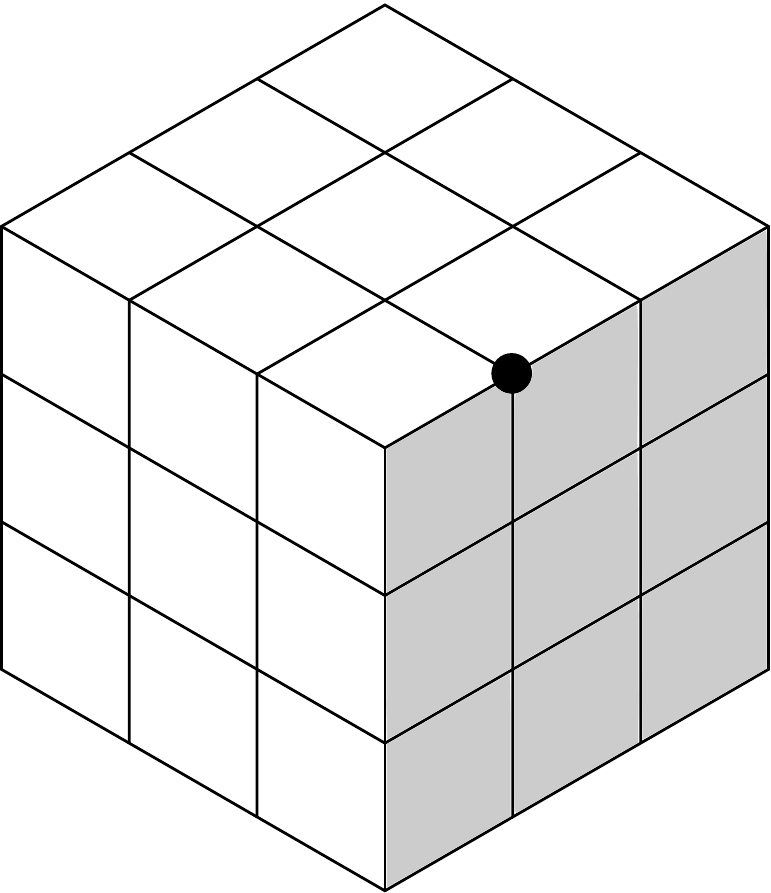}
\end{minipage}
\begin{minipage}{.1\textwidth}
\be
\xrightarrow{\; g_{ \text{even} }   \;} \nonumber
\ee
\end{minipage}
\begin{minipage}{.3\textwidth}
  \centering
  \includegraphics[width=\textwidth]{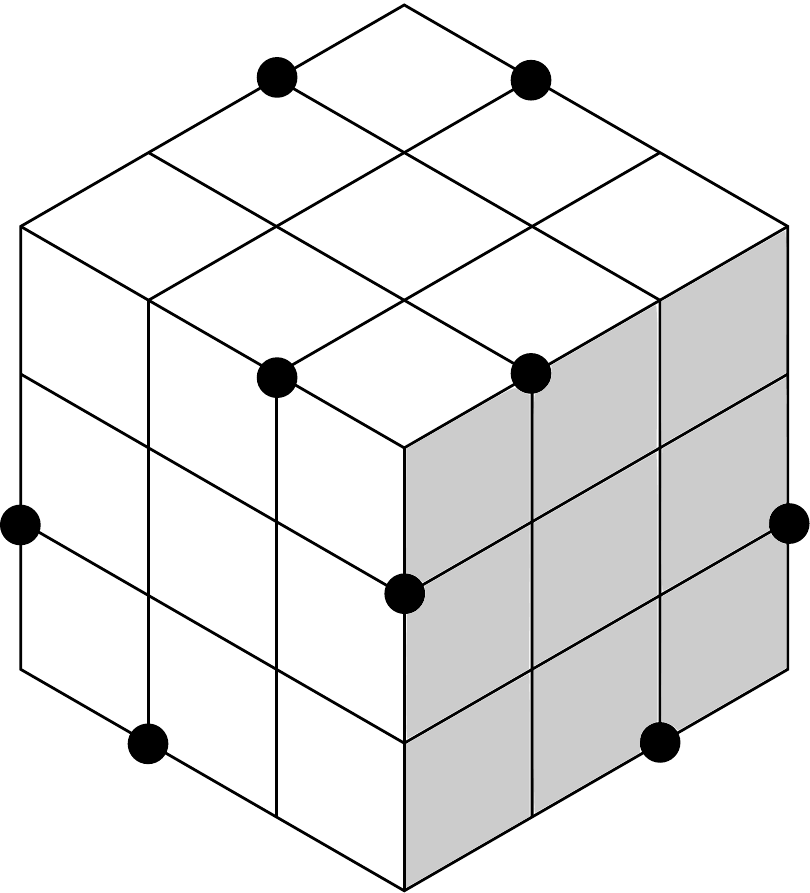}
\end{minipage}
\caption{Points generated under the action of even elements of $T_d$}
\label{fig:T_element}
\end{figure}

Although the Cartesian method was used above specifically to study the $B=20$ $T_d$-symmetric Skyrmion, it can be generalized to other Skyrmions with the same symmetry. The first question of the generalization is whether there is a modification of the F-R constraints, as this would have a great impact on the allowed states. This can be resolved with some symmetry arguments together with the help of the cubic grid. The F-R constraints only involve the 12 even elements (rotations) of the $T_d$ point group ($g_{\text{even}} \in T \subset T_d$). A set of 12 points can be generated by acting with the even elements on a generic point of the cubic grid (see Figure \ref{fig:T_element}).

A rational map can be constructed by taking such a set of points as zeros and a similar set of 12 points as poles. This results in a degree 12 rational map, which gives a $B=12$ Skyrmion upon relaxation. Rational maps of higher degree can be constructed similarly, by using more such sets of 12 points. The degree of a rational map, constructed by this method, is a multiple of 12; the baryon number of the resulting Skyrmion has the general form $B=12n$, $n \in \mathbb{N}$, when $n$ sets of points are used for the zeros and $n$ sets for the poles. The points on the vertices of the grid are special, because only four points are generated by the action of the even elements, $g_{\text{even}}$. If there are zeros on four tetrahedrally-related vertices, there need to be poles on the other four. This modifies the baryon number to $B=12n + 4m$, when zeros and poles of multiplicity $m$ on the vertices are included. The general rational map of this geometrical type has the form
\be
R_{T_d(m,n)}= \left(\frac{p_+}{p_-}\right)^m \left(\frac{\prod_{i=1}^{n} p_i}{\prod_{i=1}^{n} q_i}\right) \,.
\label{eqn:Tdgen_rat}
\ee
Here, $p_+$ and $p_-$ are the Klein polynomials given in (\ref{eqn:Klein}); each $p_i$ and $q_i$ is a degree 12 polynomial constructed from a generic set of 12 $T$-related points. $p_i$ and $q_i$ are Klein polynomials having the following general forms \cite{FLM},
\be
p_i = \frac{1}{1+a_i}(a_i p_+^3 + p_-^3) \, , \qquad q_i = \frac{1}{1+b_i} ( b_i p_+^3 + p_-^3 ) \,,
\ee
where $a_i$ and $b_i$ are mutually distinct constants.

Under the $\frac{2\pi}{3}$-rotation associated to the operator $e^{i\frac{2\pi}{3\sqrt{3}}(\hat{L}_1+\hat{L}_2+\hat{L}_3)}$, the rational map (\ref{eqn:Tdgen_rat}) picks up a phase factor which depends on the value $m$, but not $n$,
\be
R_{T_d(m,n)} \rightarrow e^{i\frac{2m\pi}{3}} R_{T_d(m,n)} \, .
\ee
This implies that the corresponding F-R constraint is dictated by the multiplicity $m$ of the vertex points; it is
\be
e^{i \frac{2 \pi}{3 \sqrt{3}} (\hat{L}_1 + \hat{L}_2 + \hat{L}_3)} e^{i \frac{2m\pi}{3} \hat{K}_3} \left | \Psi \right \rangle = (-1)^{\mathcal{N}_1} \left | \Psi \right \rangle \,.
\label{eqn:FRmod}
\ee
\noindent Note that $m$ appears in the operator on the left hand side. $\mathcal{N}_1$ can be calculated using the formula \cite{Krusch}
\be
\mathcal{N} = \frac{B}{2 \pi} (B \alpha - \beta) \, , \label{eqn:FRsign}
\ee
where $B=12n+4m$ is the baryon number, and $\alpha=\frac{2\pi}{3}$ and $\beta=\frac{2m\pi}{3}$ are the angles of rotation in space and isospace occurring in (\ref{eqn:FRmod}). So 
\begin{align}
\mathcal{N}_1 &= \frac{(12n+4m)}{2\pi}\left((12n+4m)\frac{2 \pi}{3} - \frac{2 m \pi}{3}\right) \nonumber \\
&= \frac{1}{3}(12n+4m)(12n+4m-m) \nonumber \\
&= 4(3n+m)(4n+m) \, . \label{eqn:Tdsign}
\end{align}
$\mathcal{N}_1$ is always a multiple of $4$, hence $(-1)^{\mathcal{N}_1} = 1$.

The rational map (\ref{eqn:Tdgen_rat}) is invariant under the $\pi$-rotation associated to the operator $e^{i\pi\hat{L}_3}$,
\be
R_{T_d(m,n)} \rightarrow R_{T_d(m,n)} \, ,
\ee
hence the second F-R constraint is not modified and is independent of $m$ and $n$; it is
\be
e^{i \pi \hat{L}_3} \left | \Psi \right \rangle = (-1)^{\mathcal{N}_2} \left | \Psi \right \rangle \, .
\label{eqn:TdFR2mod}
\ee
As $\alpha=\pi$ and $\beta=0$, formula (\ref{eqn:FRsign}) gives
\begin{align}
\mathcal{N}_2 &= \frac{(12n+4m)}{2\pi}\left((12n+4m)\pi - 0 \right) \nonumber \\
&= (6n+2m)(12n+4m) \nonumber \\
&= 8(3n+m)^2 \, , \label{eqn:Tdsign_2}
\end{align}
which is again even, so $(-1)^{\mathcal{N}_2} = 1$. The conclusion is that all the $T_d$-symmetric rational maps constructed from the grid have positive F-R signs for both F-R constraints, but the first constraint equation depends on $m$.

\begin{figure}[ht]
\noindent\begin{minipage}{.25\textwidth}
  \centering
  \includegraphics[width=\textwidth]{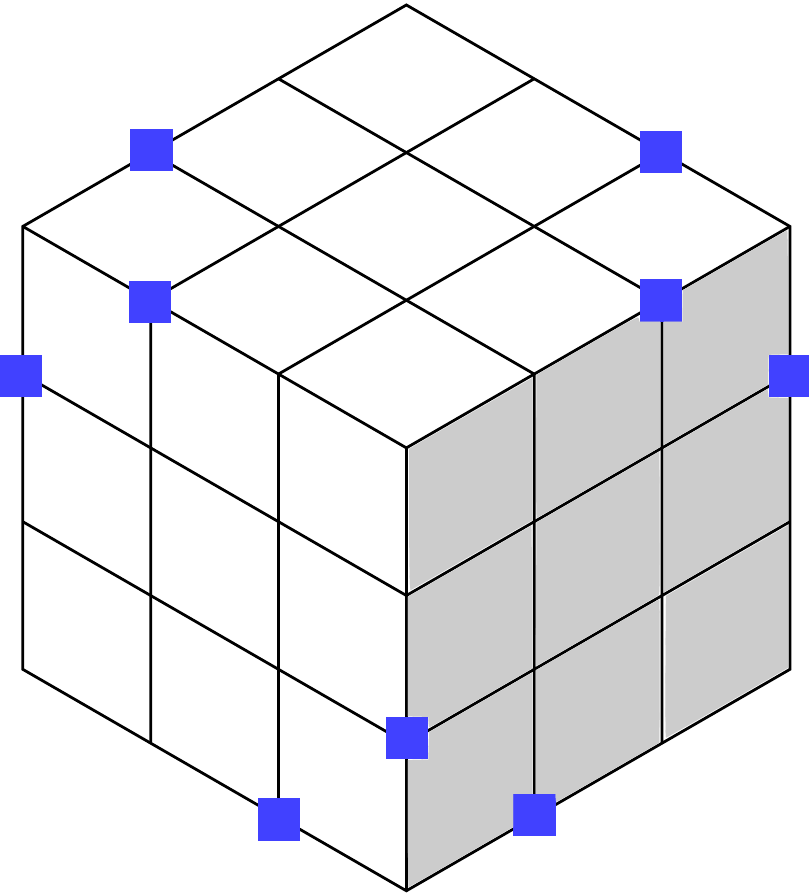}
\end{minipage}
\begin{minipage}{.1\textwidth}
\be
\xrightarrow{\; \text{  \large{$I_{\textrm{s}}$}  } \;} \nonumber
\ee
\end{minipage}
\begin{minipage}{.25\textwidth}
  \centering
  \includegraphics[width=\textwidth]{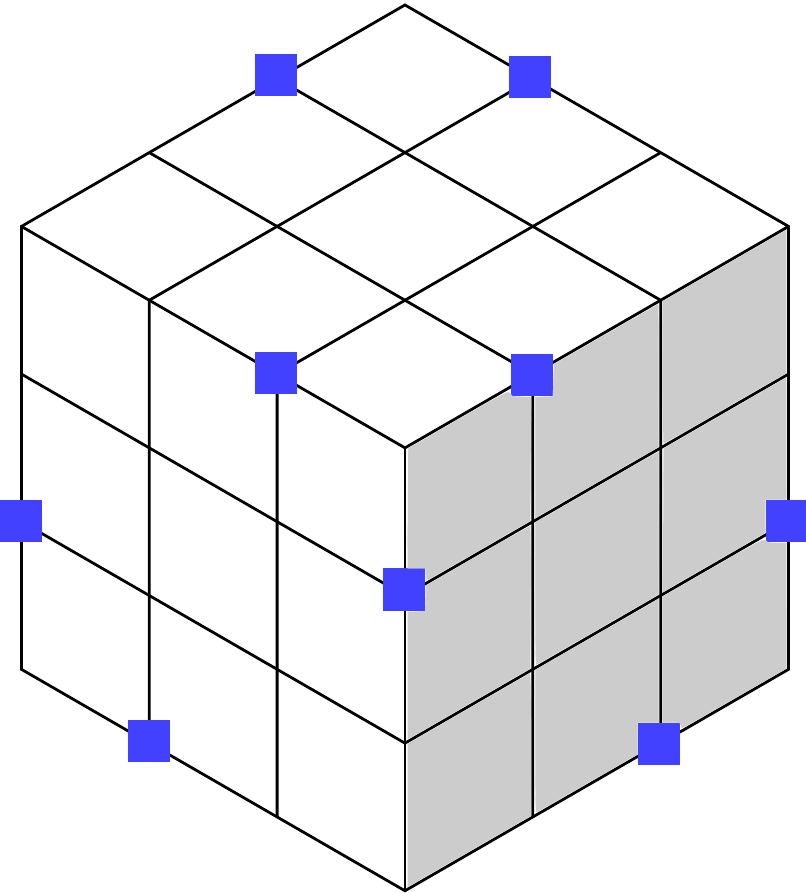}
\end{minipage}
\begin{minipage}{.1\textwidth}
\be
\xrightarrow{\text{ \large{$\sigma_{\textrm{iso}}$} }} \nonumber
\ee
\end{minipage}
\begin{minipage}{.25\textwidth}
  \centering
  \includegraphics[width=\textwidth]{T_elements_mI.pdf}
\end{minipage}
\caption{$T_d$-symmetric rational map under $I_{\textrm{s}}$ followed by $\sigma_{\textrm{iso}}$}
\label{fig:Tdinv}
\end{figure}

Using the cubic grid, one can also show that all $T_d$-symmetric rational maps have the same parity operator. Figure \ref{fig:Tdinv} shows the effect of applying $I_{\textrm{s}}$ followed by $\sigma_{\textrm{iso}}$ on a set of 12 generic points generated by the even elements of $T_d$. The effect is the same as for the $B=20$ $T_d$-symmetric Skyrmion; it is a $\frac{\pi}{2}$-rotation about the $x_3$-axis combined with a $\pi$-rotation about the $X$-axis, that is,
\be
R(\textrm{z}) \xrightarrow{\quad \text{\large{$I_{ \textrm{s}}$ , $\sigma_{\textrm{iso}}$ }} \quad} \frac{1}{\overline{R}(-\frac{1}{\bar{\textrm{z}}})} = \frac{1}{R(i\textrm{z})} \xrightarrow{\quad \text{\large{$C_{2 \, \textrm{iso}}$}} \quad} -\frac{1}{R(i\textrm{z})} \,.
\ee
The parity operator is therefore the same as before, $\hat{P}=e^{i \frac{\pi}{2} \hat{L}_3} e^{i \pi \hat{K}_2}$. The trichotomy of F-R constraints allows us to classify the quantum states of $T_d$-symmetric Skyrmions, constructed from the cubic grid, into three classes. The classification is discussed in the following subsection. 

\subsection{Classification of $T_d$-symmetric quantum states}

The isospin operator in the F-R constraint (\ref{eqn:FRmod}) depends on the value $m$. The polynomials $g_\omega$ and $g_\omega^2$ pick up different phase factors under this operator (see Table \ref{tab:FRm}), and this leads to three classes of allowed states.

\begin{table}[h]
\begin{center}
\begin{tabular}{ | c | c | c | c | }
\hline
Polynomials & $m=0 \pmod 3$ & $m=1 \pmod 3$ & $m=2 \pmod 3$ \\ \hline
$g_\omega$ & $g_{\omega}$ & $\omega \, g_{\omega}$ & $\omega^2 \, g_{\omega}$ \\ \hline
$g_{\omega^2}$ & $g_{\omega^2}$ &$\omega^2 g_{\omega^2}$ & $\omega \, g_{\omega^2}$ \\
\hline 
\end{tabular}
\end{center}
\caption{Transformation of $g_{\omega}$ and $g_{\omega^2}$ under the isospin operator in (\ref{eqn:FRmod})}
\label{tab:FRm}
\end{table}

\noindent For $m=0 \pmod 3$, the F-R constraints simplify to
\be
e^{i \frac{2 \pi}{3 \sqrt{3}} (\hat{L}_1 + \hat{L}_2 + \hat{L}_3)} \left | \Psi \right \rangle = \left | \Psi \right \rangle \,, \qquad e^{i \pi \hat{L}_3} \left | \Psi \right \rangle = \left | \Psi \right \rangle \,.
\ee
They are the same as (\ref{eqn:B20_FRK0}), which are the F-R constraints for the $B=20$ $T_d$-symmetric Skyrmion with $K=0$. The value of isospin is not constrained, and the allowed values of spin are those given in Table \ref{tab:quantum} with $K=0$, i.e. $L=0,3,4$. States of spin $L$ and isospin $K$ form an isospin ($2K+1$)-plet; the example of states with $L=3$ and $K=1$ is shown in Table \ref{tab:m0L3K1}. 

\begin{table}[ht]
\begin{center}
\begin{tabular}{ | c | c | p{5cm} | p{3cm} | }
\hline
Spin, $L^{P}$ & Isospin, $K$ & Angular momentum basis & Cartesian basis\\ \hline
$3^{+}$ & $1$ & $\left | 3 , \pm 2 \right \rangle_{-}  \otimes \left | 1 , 0 \right \rangle $ & $f_{3}g_{1}$ \\ \hline
$3^{+}$ & $1$ & $\left | 3 , \pm 2 \right \rangle_{-}  \otimes \left( \left | 1 , 1 \right \rangle + \left | 1 , -1 \right \rangle \right)$ & $f_{3}(g_{\omega}-g_{\omega^2})$ \\ \hline
$3^{-}$ & $1$ & $\left | 3 , \pm 2 \right \rangle_{-}  \otimes \left(\left | 1 , 1 \right \rangle - \left | 1 , -1 \right \rangle \right)$ & $f_{3}(g_{\omega}+g_{\omega^2})$ \\
\hline 
\end{tabular}
\end{center}
\caption{$L=3$ and $K=1$ triplet for $m=0 \pmod{3}$}
\label{tab:m0L3K1}
\end{table}

For $m=1 \pmod 3$, the F-R constraints are the same as those for the $B=20$ $T_d$-symmetric Skyrmion,
\be
e^{i \frac{2 \pi}{3 \sqrt{3}} (\hat{L}_1 + \hat{L}_2 + \hat{L}_3)} e^{i \frac{2\pi}{3} \hat{K}_3} \left | \Psi \right \rangle = \left | \Psi \right \rangle \,, \qquad e^{i \pi \hat{L}_3} \left | \Psi \right \rangle = \left | \Psi \right \rangle \,.
\ee
The allowed states are identical to the states given in Table \ref{tab:quantum}.

For $m=2 \pmod 3$, the states in Table \ref{tab:quantum} that involve $g_\omega$ and $g_{\omega^2}$ no longer satisfy the F-R constraint (\ref{eqn:FRmod}). The roles of $g_\omega$ and $g_{\omega^2}$ are exchanged; for example, the $L=2$ and $K=1$ positive parity state $F^+_{2,1}=f_\omega g_{\omega^2} - f_{\omega^2} g_\omega$ for $m=1$ is replaced by $F^+_{2,1}=f_\omega g_{\omega} - f_{\omega^2} g_{\omega^2}$ for $m=2$. The list of states for $m=2$ can be obtained from Table \ref{tab:quantum} by exchanging $g_\omega$ by $g_{\omega^2}$.

\subsection{$B=56$ $T_d$-symmetric Skyrmion}

\begin{figure}[ht]
\centering
  \includegraphics[width=0.3\textwidth]{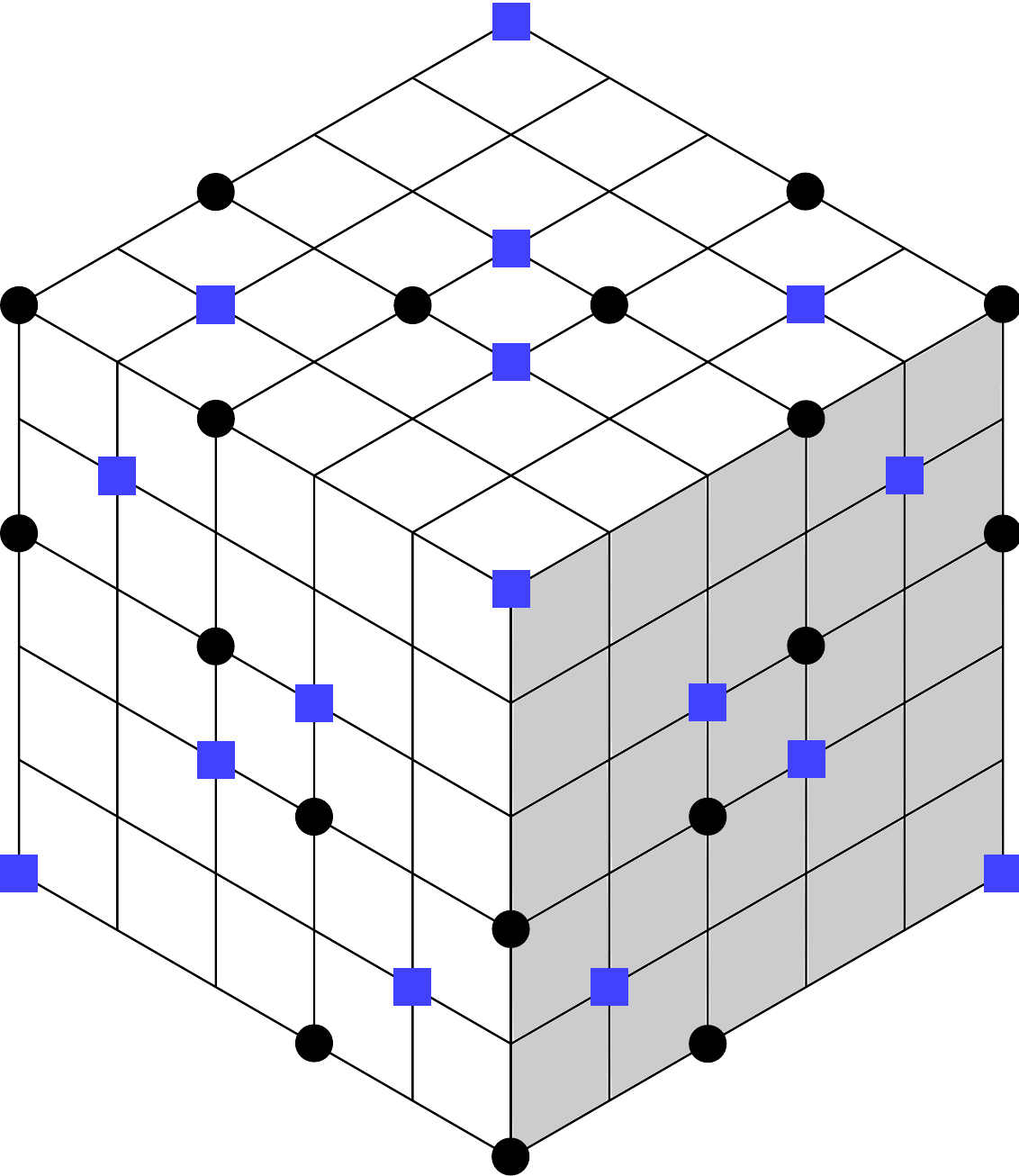}
\caption{Outer rational map for the $B=56$ $T_d$-symmetric Skyrmion}
\label{fig:cubic_grid_56}
\end{figure}

An application of the classification is to the quantum states of a $T_d$-symmetric Skyrmion with baryon number $B=56$. This Skyrmion is found by using a double-layer rational map ansatz. The first layer of the cubic grid is left out, and the inner, degree 28 rational map comes from the second layer of the cubic grid (see Figure \ref{fig:cube_grid}). The outer, degree 28 rational map is constructed from the third layer of the cubic grid; the points used in the construction are shown in Figure \ref{fig:cubic_grid_56}.

\begin{figure}[ht]
\centering
  \includegraphics[width=0.3\textwidth]{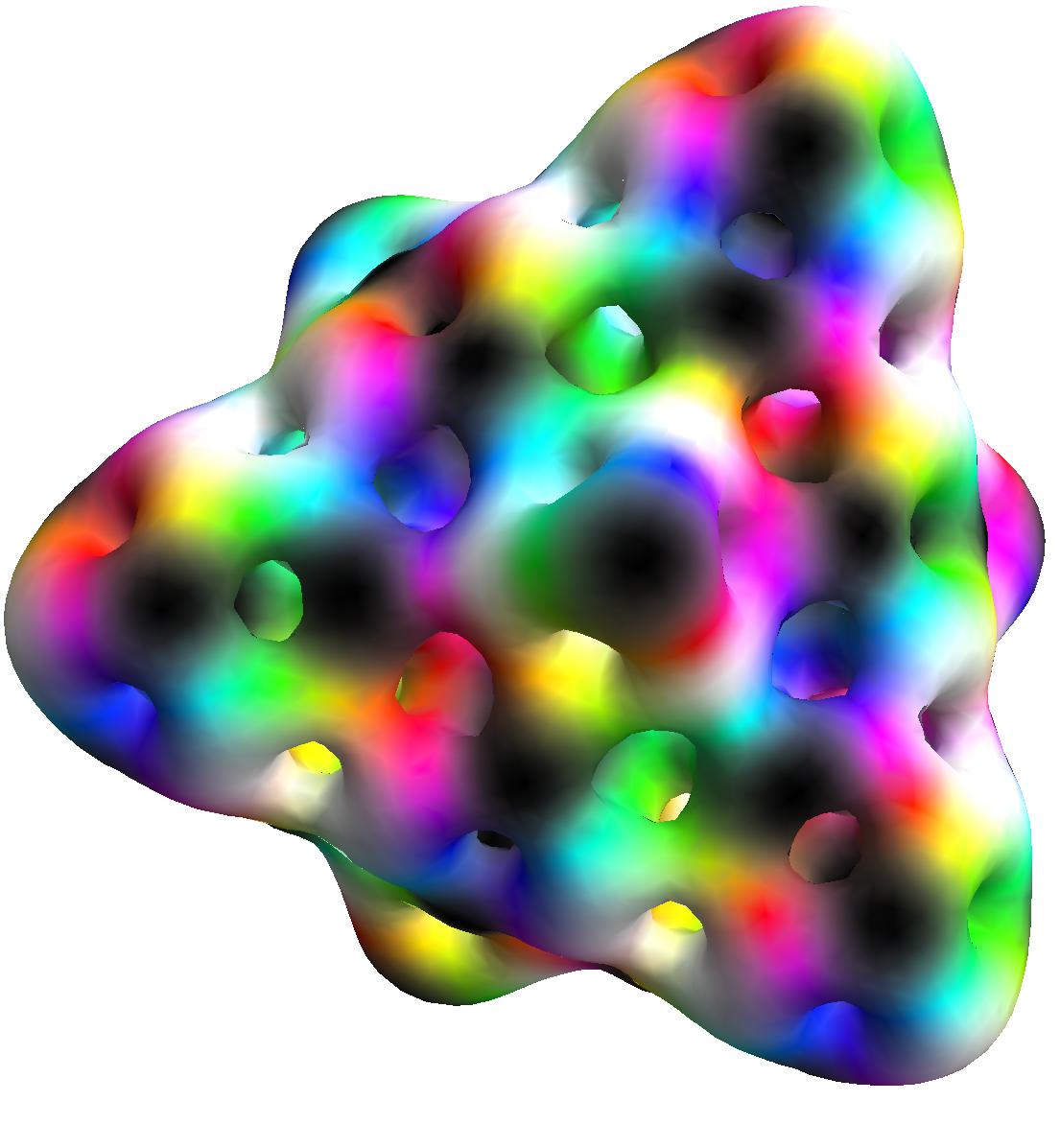}
\caption{$B=56$ Skyrmion with $T_d$ symmetry}
\label{fig:B_56}
\end{figure}

\noindent The outer rational map is
\be
R_{56,T_d} = \left( \frac{1+d_2}{1+d_1} \right) \frac{p_+}{p_-} 
\left(\frac{d_1 p_+^3 + p_-^3}{d_2 p_+^3 + p_-^3}\right) \left( \frac{d_3 p_+^3 + p_-^3}{p_+^3 + d_3 p_-^3} \right) \, ,
\ee
\noindent where $d_1=2.715$, $d_2=-22.253$ and $d_3=-0.670$. For a more detailed discussion of the interpretation of such constants, see ref.\cite{FLM}. The $B=56$ Skyrmion obtained by relaxing these rational maps preserves the $T_d$ symmetry, and it is a higher baryon number analogue of the $B=20$ Skyrmion. The Skyrmion is shown in Figure \ref{fig:B_56}. It is worth noting that if one uses a triple-layer rational map ansatz, including the first layer of the cubic grid as the innermost, degree 4 rational map, one finds a $B=60$ $T_d$-symmetric Skyrmion that looks similar. 

The primary colours, red, green and blue, are cyclically permuted when a $\frac{2\pi}{3}$-rotation is applied about a vertex, and the corresponding isorotation is a $\frac{2\pi}{3}$-rotation about the $Z$-axis. It implies that the $B=56$ Skyrmion has the same $T_d$ symmetry realization as the $B=20$ Skyrmion. They therefore have the same set of F-R constraints with $m=1$ and the same F-R signs. The $B=56$ Skyrmion has the same parity operator as the $B=20$ Skyrmion too, which can be shown using the pictorial method. They therefore have the same set of allowed spin/isospin states. The states only differ in their excitation energies. The moments of inertia of the $B=56$ Skyrmion are larger than those of the $B=20$ Skyrmion because of its size, and the excitation energy of each state is lower.

\section{$O_h$-symmetric Skyrmions}

\begin{figure}[ht]
\centering
\noindent\begin{minipage}{.3\textwidth}
  \centering
  \includegraphics[width=\textwidth]{T_elements_1.pdf}
\end{minipage}
\begin{minipage}{.1\textwidth}
\be
\xrightarrow{\; g_{ \text{even} }   \;} \nonumber
\ee
\end{minipage}
\begin{minipage}{.3\textwidth}
  \centering
  \includegraphics[width=\textwidth]{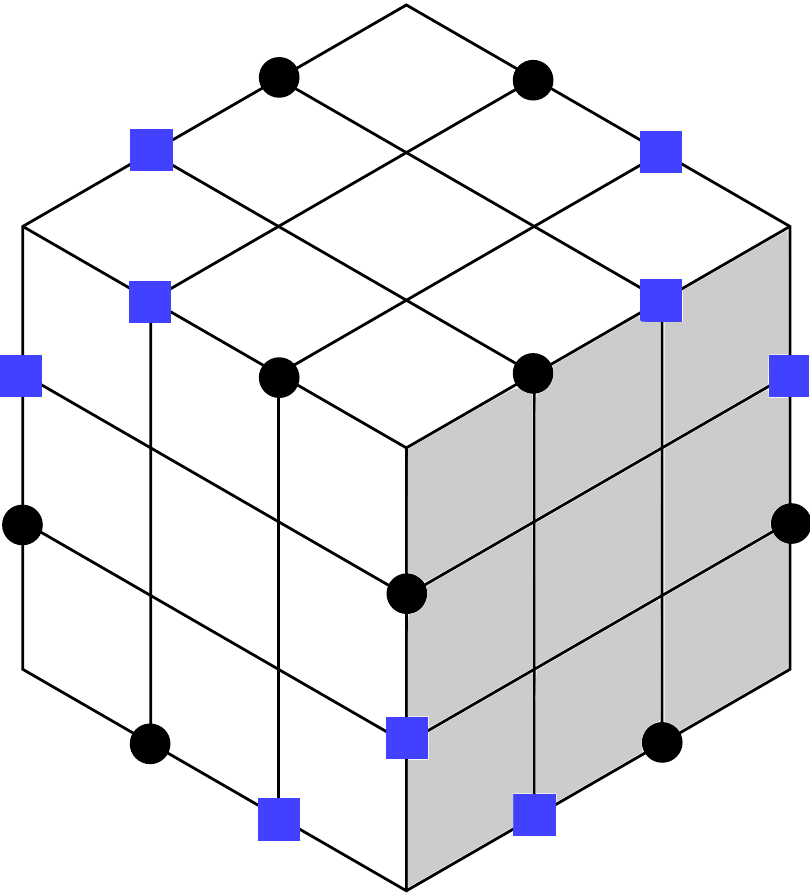}
\end{minipage}
\caption{Points generated under the action of even elements of $O_h$}
\label{fig:O_element}
\end{figure}

The $T_d$ symmetry group, which we have been studying, is a normal subgroup of the cubic group $O_h$, and the $O_h$ group is the full symmetry group of the cubic grid. Thus, it is natural to extend our techniques to $O_h$. $O_h$ ($48$ elements) is twice as big as $T_d$ ($24$ elements). This imposes more constraints on the states, and fewer states are allowed. Using the $B=32$ $O_h$-symmetric Skyrmion as an example, the corresponding F-R constraints are
\be
e^{i \frac{2 \pi}{3 \sqrt{3}} (\hat{L}_1 + \hat{L}_2 + \hat{L}_3)} e^{i \frac{2\pi}{3} \hat{K}_3} \left | \Psi \right \rangle = \left | \Psi \right \rangle \, , \qquad
e^{i \frac{\pi}{2} \hat{L}_3} e^{i \pi \hat{K}_1} \left | \Psi \right \rangle = \left | \Psi \right \rangle \, .
\label{eqn:OhFR}
\ee
24 points will be generated from a generic point under the action of the 24 even elements of $O_h$, hence one may think that the baryon number is of the form $B=24n$, $n \in \mathbb{N}$; however, this is incorrect because of the second F-R constraint. The operators of the constraint turn zeros into poles, and vice versa. The set of $24$ points consists of $12$ zeros and $12$ poles; therefore, the baryon number is of the form $B=12n$, when $n$ sets are included. Including the points on the vertices, we obtain the same formula as in the $T_d$ case, $B=12n+4m$, $n,m \in \mathbb{N}$. The only difference between the $T_d$ and $O_h$ symmetry is the extra constraint that relates the zeros and poles (see Figure \ref{fig:O_element}).

\begin{figure}[ht]
\noindent\begin{minipage}{.25\textwidth}
  \centering
  \includegraphics[width=\textwidth]{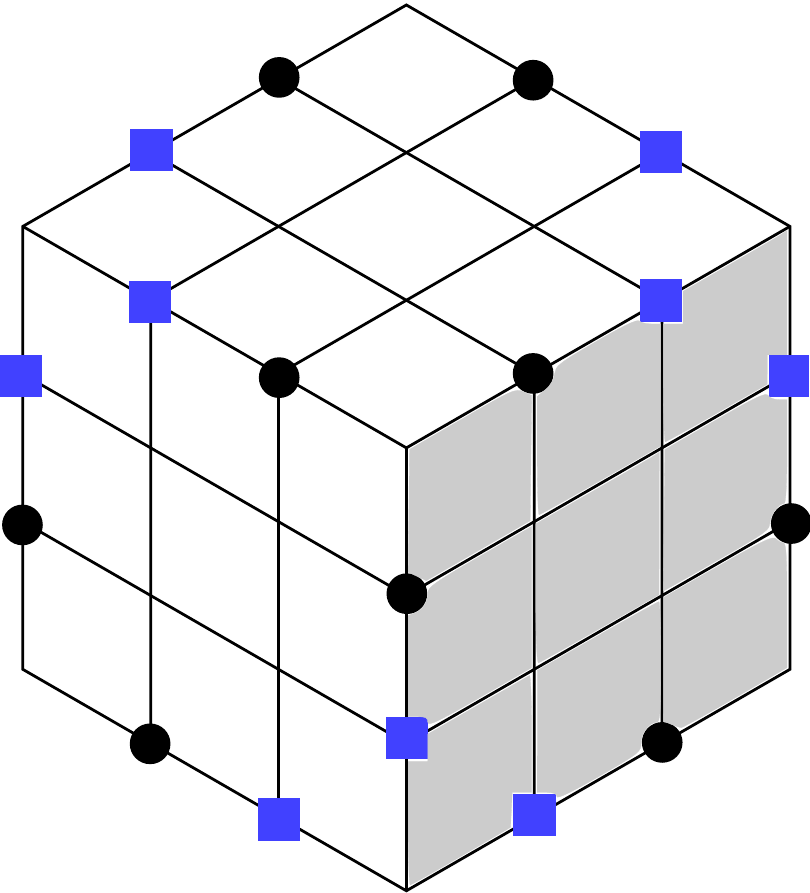}
\end{minipage}
\begin{minipage}{.1\textwidth}
\be
\xrightarrow{\; \text{  \large{$I_{\textrm{s}}$}  } \;} \nonumber
\ee
\end{minipage}
\begin{minipage}{.25\textwidth}
  \centering
  \includegraphics[width=\textwidth]{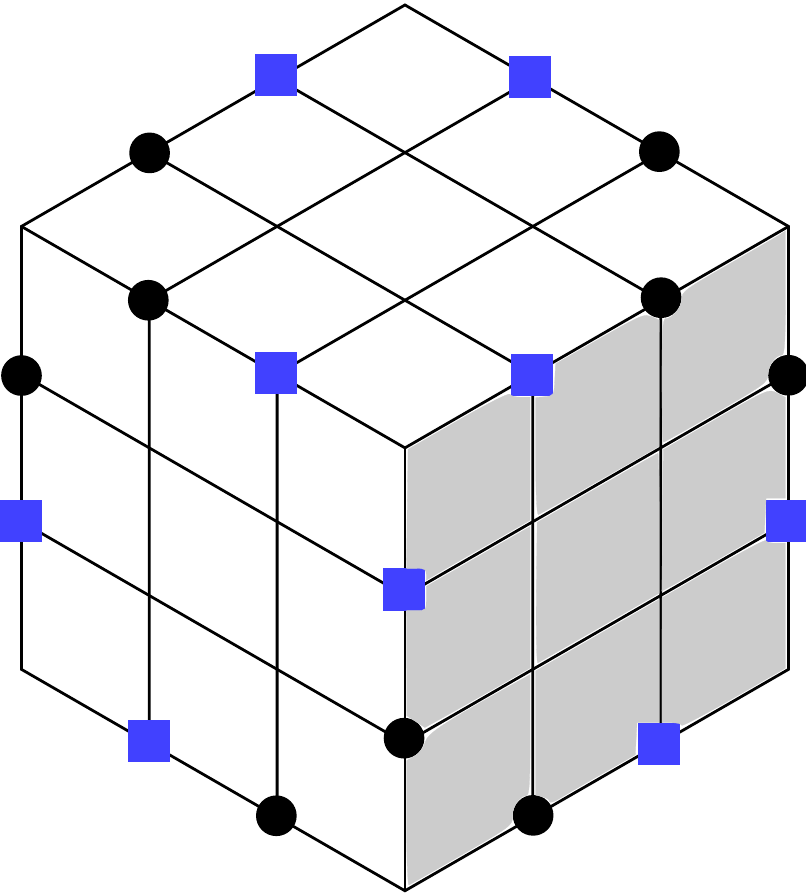}
\end{minipage}
\begin{minipage}{.1\textwidth}
\be
\xrightarrow{\text{ \large{$\sigma_{\textrm{iso}}$} }} \nonumber
\ee
\end{minipage}
\begin{minipage}{.25\textwidth}
  \centering
  \includegraphics[width=\textwidth]{Oh_general_grid.pdf}
\end{minipage}
\caption{$O_h$-symmetric rational map under $I_{\textrm{s}}$ followed by $\sigma_{\textrm{iso}}$}
\label{fig:Oh_inv}
\end{figure}

To derive the generalized F-R constraints for $O_h$-symmetric Skyrmions, constructed from the cubic grid, we consider a general $O_h$-symmetric rational map given by
\be
R_{O_h(m,n)}=\left( \frac{p_+}{p_-} \right)^m \prod_{i=1}^n \left(\frac{a_i p_+^3 + p_-^3}{p_+^3 + a_i p_-^3}\right) \,,
\label{eqn:Ohgen_rat}
\ee
where $a_i$ is constant. The $T_d$-symmetric rational map (\ref{eqn:Tdgen_rat}) can be specialized to an $O_h$-symmetric rational map by imposing the extra conditions $b_i = \rfrac{1}{a_i}$,
\begin{align}
R_{T_d(m,n)}\Bigg|_{b_i=\frac{1}{a_i}}&= \left(\frac{p_+}{p_-}\right)^m \left(\frac{\prod_{i=1}^{n} p_i}{\prod_{i=1}^{n} q_i}\right)\Bigg|_{b_i=\frac{1}{a_i}} \nonumber \\
&=\left(\frac{p_+}{p_-}\right)^m \left(\frac{\prod_{i=1}^{n} \frac{1}{1+a_i}(a_i p_+^3 + p_-^3)}{\prod_{i=1}^{n} \frac{1}{1+b_i}(b_i p_+^3 + p_-^3)}\right)\Bigg|_{b_i=\frac{1}{a_i}} \nonumber \\
&= \left(\frac{p_+}{p_-}\right)^m \prod_{i=1}^{n} \left(\frac{a_i p_+^3 + p_-^3}{p_+^3 + a_i p_-^3} \right) \nonumber \\
&= R_{O_h(m,n)} \, .
\end{align}
\indent The $O_h$-symmetric rational map transforms under the rotation associated to the operator $e^{i \frac{2 \pi}{3 \sqrt{3}} (\hat{L}_1 + \hat{L}_2 + \hat{L}_3)}$ as
\be
R_{O_h(m,n)} \rightarrow e^{i\frac{2m\pi}{3}}R_{O_h(m,n)}
\ee
and under the operator $e^{i\frac{\pi}{2}\hat{L}_3}$ as
\be
R_{O_h(m,n)} \rightarrow \frac{1}{R_{O_h(m,n)}} \,.
\ee
This leads to the following generalized $O_h$ F-R constraints,
\be
e^{i \frac{2 \pi}{3 \sqrt{3}} (\hat{L}_1 + \hat{L}_2 + \hat{L}_3)} e^{i \frac{2m\pi}{3} \hat{K}_3} \left | \Psi \right \rangle = (-1)^{\mathcal{N}_1}\left | \Psi \right \rangle \, , \qquad
e^{i \frac{\pi}{2} \hat{L}_3} e^{i \pi \hat{K}_1} \left | \Psi \right \rangle = (-1)^{\mathcal{N}_2}\left | \Psi \right \rangle \, .
\label{eqn:OhFR}
\ee
The first F-R constraint is the same as the $T_d$-symmetric case, while the second F-R constraint is a ``square root'' of the $T_d$ F-R constraint (\ref{eqn:TdFR2mod}).
$\mathcal{N}_1$ is identical to (\ref{eqn:Tdsign}) and is even; $\mathcal{N}_2$ can be calculated using (\ref{eqn:FRsign}) with $B=12n+4m$, $\alpha=\frac{\pi}{2}$ and $\beta=\pi$,
\begin{align}
\mathcal{N}_2 &= \frac{(12n+4m)}{2\pi}\left((12n+4m)\frac{\pi}{2} - \pi \right) \nonumber \\
&= (6n+2m)(6n+2m - 1) \nonumber \\
&= 2(3n+m)(6n+2m-1) \, , 
\label{eqn:Ohsign_2}
\end{align}
\noindent which is still even, hence $(-1)^{\mathcal{N}_2}=1$. The operators of the second F-R constraint transform the Cartesian coordinates as
\begin{align}
x &\rightarrow y & &, & y &\rightarrow -x & &, & \ z &\rightarrow z & \, &; \\
X &\rightarrow X & &, & Y &\rightarrow -Y & &, & \ Z &\rightarrow -Z & \, &.
\end{align}
The parity operator is different from the $T_d$ case and can be extracted with the pictorial method as shown in Figure \ref{fig:Oh_inv}.
\noindent The rational map $R(\mathrm{z})$ returns to its original form after the transformation $I_{\textrm{s}}$ and $\sigma_{\textrm{iso}}$. Including the isorotation $C_{2 \, \textrm{iso}}$, one finds that $R(\mathrm{z})$ transforms to $-R(\mathrm{z})$. The parity operator is therefore $\hat{P}=e^{i \pi \hat{K}_3}$.

\begin{table}[ht]
\begin{center}
\begin{tabular}{ | c | c | c | c | c | c | }
\hline
Degree & & $T$-invariant polynomials & $1^{\text{st}}$ F-R & $2^{\text{nd}}$ F-R & Parity\\ \hline
$2$ & $f_2$ & $x^2 + y^2 + z^2$ & $f_2$ & $f_2$ & $+1$ \\ \hline
$3$ & $f_3$ & $xyz$ & $f_3$ & $-f_3$ & $+1$ \\ \hline
$4$ & $f_4$ & $x^4 + y^4 + z^4$ & $f_4$ & $f_4$ & $+1$ \\ \hline
$6$ & $f_6$ & $x^4(y^2-z^2) + y^4(z^2-x^2) + z^4(x^2-y^2)$ & $f_6$ & $-f_6$ & $+1$ \\ \hline
$1$ & $g_1$ & $Z$ & $g_1$ & $-g_1$ & $+1$ \\ \hline
$2$ & $g_2$ & $X^2 + Y^2$ & $g_2$ & $g_2$ & $+1$ \\ \hline
$3$ & $g_{3-}$ & $X(X^2-3Y^2)$ & $g_{3-}$ & $g_{3-}$ & $-1$ \\ \hline
$3$ & $g_{3+}$ & $Y(3X^2-Y^2)$ & $g_{3+}$ & $-g_{3+}$ & $-1$ \\ \hline
$2$ & $f_{\omega}$ & $\omega^2(x^2 + \omega^2 y^2 + \omega z^2)$ & $\omega \, f_{\omega}$ & $f_{\omega^2}$ & $+1$ \\ \hline
$2$ & $f_{\omega^2}$ & $\omega(x^2+\omega y^2 +\omega^2 z^2)$ & $\omega^2 f_{\omega^2}$ & $f_{\omega}$ & $+1$ \\ \hline
$1$ & $g_{\omega}$ & $(X+iY)$ & $\omega^m g_{\omega}$ & $g_{\omega^2}$ & $-1$ \\ \hline
$1$ & $g_{\omega^2}$ & $(X-iY)$ & $(\omega^2)^m g_{\omega^2}$ & $g_{\omega}$ & $-1$ \\
\hline 
\end{tabular}
\end{center}
\caption{$T$-invariant polynomials under $O_h$ F-R operators}
\label{tab:cubic}
\end{table}

The $O_h$-symmetric states can be found by using the $T$-invariant polynomials. Table \ref{tab:cubic} shows the effect on the $T$-invariant polynomials of the operators of the $O_h$ F-R constraints (\ref{eqn:OhFR}). The polynomials $f_3$, $f_6$, $g_1$ and $g_{3+}$ are no longer invariant. They gain a minus sign under the operator occurring in the second constraint. We have to combine an even number of these polynomials to form $O_h$-invariants, for example, $f_3^2$ and $g_1g_{3+}$. In Table \ref{tab:quantum}, some of the $T_d$-symmetric states consist of an odd number of these polynomials; these are not allowed as states for an $O_h$-symmetric Skyrmion.

The allowed quantum states of $O_h$-symmetric Skyrmions belong to a subset of the $T_d$-symmetric states for each value of $m$. For example, the $3^-$ state is forbidden while the $4^+$ state is allowed for a quantized Skyrmion with $O_h$ symmetry. 

\section{Conclusions}

In this paper, we systematically studied the semi-classical quantization of the families of $T_d$- and $O_h$-symmetric Skyrmions constructed using the cubic grid method. A Cartesian method of solving the Finkelstein-Rubinstein (F-R) constraints was developed and applied to $T_d$-symmetric Skyrmions and $O_h$-symmetric Skyrmions. The action of the $T_d$ and $O_h$ groups on the quantum states of Skyrmions is more transparent in the Cartesian coordinates, and the F-R constraints are easier to solve.

The $T_d$- and $O_h$-symmetric rational maps constructed from the cubic grid have similar structures, and they lead to Skyrmions with baryon number a multiple of four. The symmetries of the Skyrmions are realized in almost the same way. The F-R constraints of the Skyrmions differ in the contribution of the vertex points of the cubic grid to the rational maps, and this difference is characterized by an integer $m \pmod{3}$. We classified all quantum states of the Skyrmions into three classes depending on this integer $m$. All the $T_d$- and $O_h$-symmetric Skyrmions in each class have the same set of spin and isospin quantum states.

We applied this classification to the quantum states of the $B=20$ and newly found $B=56$ $T_d$-symmetric Skyrmions. The class which these Skyrmions belong to was identified easily by looking at the F-R constraints (they have $m=1$), and the quantum states for these Skyrmions are listed in Table 4. The energies of the states of the $B=20$ Skyrmion are given in Table 5, and for $B=56$, they would be similar.

The next step would be to relate the results presented here to other studies of possibly tetrahedrally- and cubically-symmetric nuclei \cite{DGSM,DCDDPOS}; another possible step is to extend the classification method to other normal subgroups of $O_h$, for example $D_{2h}$, which is the symmetry group of another $B=20$ Skyrmion.

\clearpage

\appendix
\section{Appendix: Numerical methods}

In order to compute the energy levels of the quantized Skyrmions, the moments of inertia ($U_{ij}$,$V_{ij}$,$W_{ij}$) are needed; they are calculated numerically. First, stable Skyrmion solutions are found by applying the non-linear conjugate gradient method to the initial Skyrme field configuration, constructed from the rational map ansatz \cite{FLM,Fei}. The moments of inertia of the stable Skyrmions are then calculated. Since the Skyrme field $U(\mathbf{x})$ is $SU(2)$-valued, we cannot apply the formulae (\ref{eqn:U}), (\ref{eqn:V}) and (\ref{eqn:W}) directly. We adapt the method in \cite{BS3c} and express the Skyrme field in a 4-vector form,
\be
U= \sum_{\mu=0}^{3} \phi_{\mu} e_{\mu} \, ,
\label{eqn:fourvec}
\ee
\noindent where $\phi_{\mu} = (\sigma, \pi_j)$ and $e_\mu = ({\boldsymbol I},i \pauli_j)$, $j=1,2,3$, and we impose the constraint $\sum_{\mu=0}^{3} \phi_\mu \phi_\mu = 1$.

\noindent Substituting (\ref{eqn:fourvec}) into (\ref{eqn:U}), (\ref{eqn:V}) and (\ref{eqn:W}), the moments of inertia become
\begin{align}
U_{ij} &= \int 2 \Big( \bm{\phi}^2 \delta_{ij} - \phi_i \phi_j + (\bm{\partial} \phi_0)^2 \delta_{ij} + \bm{\phi}^2 (\bm{\partial} \phi_i \bm{\partial} \phi_j) - (\bm{\partial} \phi_0)^2 \phi_i \phi_j + \phi_0 \phi_j (\bm{\partial} \phi _0 \bm{\partial} \phi_i) \nonumber \\ & \qquad+ \phi_0 \phi_i (\bm{\partial} \phi_0 \bm{\partial} \phi_j) \Big) d^3x \,, \label{eqn:UC} \\
W_{ij} &= \int 2 \epsilon _{jlm} x_l \Big[ \epsilon_{iab} \Big( \phi_a \partial_m \phi_b + \phi_0 \partial_m \phi_a (\bm{\partial} \phi_0 \bm{\partial} \phi_b) + \phi_a \partial_m \phi_b (\bm{\partial} \phi_0)^2 - \phi_a \partial_m \phi_0 (\bm{\partial} \phi_0 \bm{\partial} \phi_b) \Big) \nonumber \\
& \qquad+ \epsilon_{abc} \phi_a \partial_m \phi_b (\bm{\partial} \phi_c \bm{\partial} \phi_i) \Big] d^3x \, , \label{eqn:WC}\\
V_{ij} &= \int 2 \epsilon _{ilm} \epsilon _{jnp} x_l x_n \Big( \partial_m \bm{\phi} \cdot \partial_p \bm{\phi} + \partial _m \phi_0 \partial _p \phi_0 + (\bm{\partial} \phi_0)^2 \partial _m \bm{\phi} \cdot \partial _p \bm{\phi} + (\bm{\partial} \bm{\phi})^2 \partial _m \phi_0 \partial _p \phi_0 \nonumber \\ 
&\qquad - (\bm{\partial} \phi_0 \bm{\partial} \bm{\phi}) \cdot \partial _p \bm{\phi} \partial _m \phi_0 - (\bm{\partial} \phi_0 \bm{\partial} \bm{\phi} ) \cdot \partial_m \bm{\phi} \partial_p \phi_0 + (\bm{\partial} \bm{\phi})^2 \partial_m \bm{\phi} \cdot \partial_p \bm{\phi} \nonumber \\
&\qquad - \partial_m \bm{\phi} \cdot (\bm{\partial} \bm{\phi} \bm{\partial} \bm{\phi}) \cdot \partial _p \bm{\phi} \Big) d^3x \, , \label{eqn:VC}
\end{align}
\noindent where $\bm{\phi}^2 = \bm{\phi} \cdot \bm{\phi} = \sum_{i=1}^{3} \phi_i \phi_i$, denotes the dot product of the pion fields, and $(\bm{\partial} \phi_i \bm{\partial} \phi_j) = \sum_{k=1}^{3} \partial_k \phi_i \partial_k \phi_j$, denotes the dot product of gradients $\bm{\partial}$.

The numerical code is written in C\texttt{++}, and several consistency tests are carried out. One test is to check the numerical values against a theoretical constraint. It is well established that the shape of the $B=2$ Skyrmion is a toroid, hence it possesses a $D_{ \infty h}$ symmetry. This implies that the moment of inertia tensors satisfy the conditions \cite{BC}

\be
U_{33} = \frac{1}{2} W_{33} = \frac{1}{4} V_{33} \, .
\label{eqn:UVWrel}
\ee

\noindent The moments of inertia of the $B=2$ toroid are calculated numerically, and the values are $U_{33}=68.67$, $V_{33}=274.59$ and $W_{33}=137.31$. Their ratios
\be
\frac{W_{33}}{U_{33}} = 1.9998 \, , \quad \frac{ V_{33} }{U_{33}} = 3.9990 \, ,
\ee
\noindent are in close agreement with the theoretical constraint.

\begin{figure}[ht]
\hspace{1.5cm}
\includegraphics[trim = 0mm 5mm 0mm 5mm, clip, width=12cm]{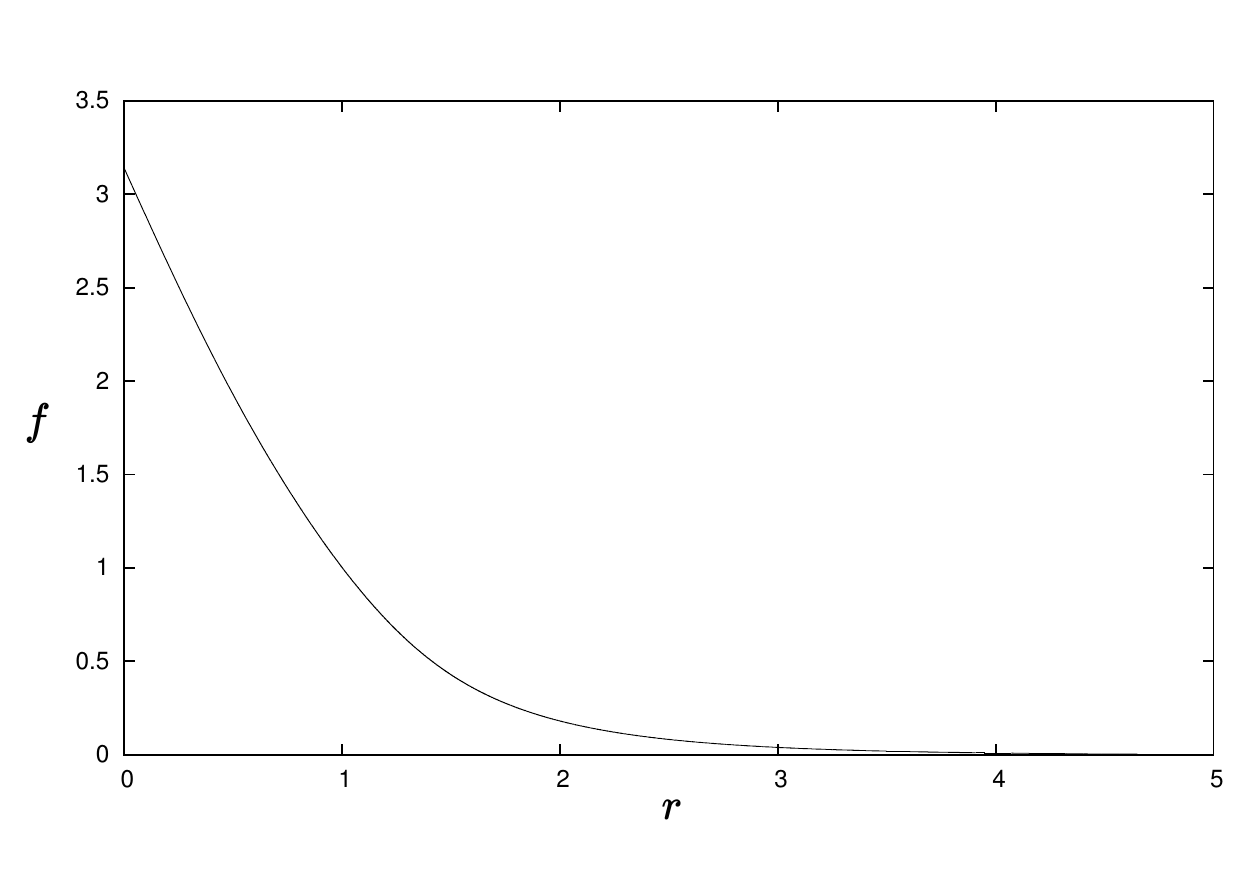}
\caption{Profile function with cut-off at $r_0=5$}
\label{fig:profile}
\end{figure}

The next test is to compare the moments of inertia calculated using two different approaches. The relatively simple form of the $B=1$ Skyrmion allows one to compute the moments of inertia numerically in two independent ways. The first approach is to use the field equation, which can be derived from the Lagrangian and is
\be
\partial_{\mu} \left( R^{\mu} + \frac{1}{4} [R^{\nu},[R_{\nu},R^{\mu}]] \right) + i \pauli_a \frac{1}{2}\Tr(-i\pauli_a U) = 0 \, ,
\label{eqn:EOM}
\ee
\noindent where the scaled pion mass is unity. The $B=1$ Skyrmion is spherically symmetric and takes the ``hedgehog'' form, which means all the pion fields point in the radial direction,
\be
U(\mathbf{x}) = \exp ( i f(r) \hat{\mathbf{x}} \cdot \pauli ) \, .
\label{eqn:hed}
\ee
\noindent Expanding out the exponential, we can identify the $\bpi$ and $\sigma$ fields,
\be
\boldsymbol{\pi} = \sin f(r) \, \hat{\mathbf{x}} \, , \quad \sigma = \cos f(r) \,,
\ee
where the radial profile function $f(r)$ is a real function satisfying the boundary conditions $f(0)=\pi$ and $f(\infty) = 0$. All the moment of inertia tensors of the $B=1$ Skyrmion are identical, because of its spherical symmetry. Therefore
\be
U_{ii} = V_{ii} = W_{ii} \, , \ i = 1 , \, 2, \, 3  ; \quad U_{ij} = V_{ij} = W_{ij} = 0 \, , \ i \neq j = 1 , \, 2 , \, 3  .
\label{eqn:UVW}
\ee
\noindent Substituting the $\bpi$ and $\sigma$ fields into the component of the inertia tensor $U_{33}$, we find in terms of the profile function $f(r)$,

\be
U_{33} = \frac{16 \pi}{3} \int^{\infty}_0 \ \sin ^2 f \left( r^2 + r^2 f'^2 + \sin ^2 f \right) dr \, ,
\label{eqn:U33}
\ee

\noindent where $f' = \frac{df}{dr}$. The profile function $f(r)$ is found using the field equation. Substituting the hedgehog ansatz (\ref{eqn:hed}) into the Skyrme field equation (\ref{eqn:EOM}), one obtains a second order nonlinear ODE for $f(r)$,
\be
(r^2 + 2 \sin ^2 f) f'' + 2rf' + \sin 2f \left( f'^2 - 1 - \frac{\sin ^2 f}{r^2} \right) - r^2 \sin f = 0 \, .
\ee
This is solved numerically using the shooting method. Since it is impossible to extend the range to $r=\infty$ numerically, and we know the function $f(r)$ falls off reasonably quickly \cite{book}, a variable cut-off $r_0 \leq 10$ is introduced. The result for $r_0=5$ is plotted in Figure \ref{fig:profile}.
\clearpage

The computed moment of inertia $U_{33}$ is
\be
U_{33}=47.60 \, .
\label{eqn:U33value}
\ee
The effect of changing the cut-off $r_0$ is presented in Table \ref{tab:cutoff}.
\begin{table}[H]
\centering
\begin{tabular}{| c | c |}
\hline
$r_0$ & $U_{33}$ \\ \hline
$4$ & $47.455$ \\ \hline
$5$ & $47.596$ \\ \hline
$6$ & $47.619$ \\ \hline
$7$ & $47.623$ \\ \hline
$8$ & $47.623$ \\ \hline
$9$ & $47.623$ \\ \hline
$10$ & $47.623$ \\
\hline
\end{tabular}
\caption{Moment of inertia with the cut-off, $r_0$}
\label{tab:cutoff}
\end{table}

The moment of inertia tensors can also be computed using the formulae (\ref{eqn:UC}), (\ref{eqn:WC}) and (\ref{eqn:VC}). A three-dimensional $B=1$ Skyrmion is prepared by relaxing the rational map ansatz with $R(\mathrm{z})=\mathrm{z}$ in a cubic box of width $8$ and lattice spacing of $0.1$ \cite{Fei}. We choose to compare calculations with similar volume; a cube of width $8$ is approximately equal to a sphere of radius $5$, by volume. The inertia tensors are presented in Table \ref{tab:B_1}. The values are in good agreement with (\ref{eqn:U33value}), and discrepancy only appears at the second decimal.

\begin{table}[ht]
\centering
\begin{tabular}{| l | l | l |}
  \hline
$U_{11}=47.561$ & $U_{12}=-3.732 \times 10^{-16}$ & $U_{13}=-3.248 \times 10^{-17}$ \\ \hline
$U_{21}=-3.367 \times 10^{-16}$ & $U_{22}=47.561$ & $U_{23}=-1.107 \times 10^{-16}$ \\ \hline
$U_{31}=-3.317 \times 10^{-17}$ & $U_{32}=-1.632 \times 10^{-16}$ & $U_{33}=47.561$ \\ \hline
\multicolumn{1}{|l}{} & \multicolumn{1}{r}{} & \\ \hline
$W_{11}=47.561$ & $W_{12}=-5.807 \times 10^{-15}$ & $W_{13}=1.035 \times 10^{-14}$ \\ \hline
$W_{21}=5.071 \times 10^{-15}$ & $W_{22}=47.561$ & $W_{23}=-2.020 \times 10^{-15}$ \\ \hline
$W_{31}=-1.019 \times 10^{-14}$ & $W_{32}=1.778 \times 10^{-15}$ & $W_{33}=47.561$ \\ \hline
\multicolumn{1}{|l}{} & \multicolumn{1}{r}{} & \\ \hline
$V_{11}=47.576$ & $V_{12}=-2.549 \times 10^{-16}$ & $V_{13}=5.652 \times 10^{-17}$ \\ \hline
$V_{21}=-5.845 \times 10^{-16}$ & $V_{22}=47.576$ & $V_{23}=-7.068 \times 10^{-17}$ \\ \hline
$V_{31}=7.812 \times 10^{-17}$ & $V_{32}=-1.088 \times 10^{-16}$ & $V_{33}=47.576$ \\
  \hline  
\end{tabular}
\caption{Moment of inertia tensors of $B=1$ Skyrmion}
\label{tab:B_1}
\end{table}

In Table \ref{tab:B_1}, the diagonal components of the tensors $U$, $V$ and $W$ differ by less than $0.05 \%$, consistent with (\ref{eqn:UVW}).

\newpage

The inertia tensors of the $B=20$ $T_d$-symmetric Skyrmion are calculated similarly and presented in Table \ref{tab:B_20}.
\begin{table}[ht]
\centering
\begin{tabular}{| l | l | l |}
  \hline
$U_{11}=758.514$ & $U_{12}=7.052 \times 10^{-14}$ & $U_{13}=-0.017$ \\ \hline
$U_{21}=7.185\times 10^{-14}$ & $U_{22}=758.508$ & $U_{23}=-2.816\times 10^{-15}$ \\ \hline
$U_{31}=-0.017$ & $U_{32}=-9.569 \times 10^{-17}$ & $U_{33}=819.931$ \\ \hline
\multicolumn{1}{|l}{} & \multicolumn{1}{r}{} & \\ \hline
$W_{11}=1.547 \times 10^{-13}$ & $W_{12}=-1.006\times 10^{-13}$ & $W_{13}=-0.002$ \\ \hline
$W_{21}=-8.020\times 10^{-13}$ & $W_{22}=3.359\times 10^{-13}$ & $W_{23}=7.862\times10^{-14}$ \\ \hline
$W_{31}=-8.220\times 10^{-13}$ & $W_{32}=-3.048\times 10^{-13}$ & $W_{33}=0.002$ \\ \hline
\multicolumn{1}{|l}{} & \multicolumn{1}{r}{} & \\ \hline
$V_{11}=12854.0$ & $V_{12}=-0.207$ & $V_{13}=5.274\times 10^{-12}$ \\ \hline
$V_{21}=-0.207$ & $V_{22}=12854.0$ & $V_{23}=3.363\times 10^{-12}$ \\ \hline
$V_{31}=5.3739\times 10^{-12}$ & $V_{32}=2.881 \times 10^{-12}$ & $V_{33}=12854.6$ \\
  \hline  
\end{tabular}
\caption{Moment of inertia tensors of $B=20$ $T_d$-symmetric Skyrmion}
\label{tab:B_20}
\end{table}

For the $B=20$ $T_d$-symmetric Skyrmion, we expect that $V_{11}=V_{22}=V_{33}$, $U_{11}=U_{22}$ and $W_{ij}=0$. The numerical values presented are consistent with the expected structure.

\section*{Acknowledgements}

P.H.C. Lau is supported by Trinity College, Cambridge. P.H.C. Lau also thanks Paul Sutcliffe for suggesting the method of calculating inertia tensors of the $B=1$ Skyrmion.

\end{document}